\documentclass[12pt]{iopart}
\usepackage{graphicx}
\expandafter\let\csname equation*\endcsname\relax
\expandafter\let\csname endequation*\endcsname\relax
\usepackage{amsmath}
\usepackage{mathabx}
\usepackage{cite}
\usepackage{lscape}
\usepackage{varwidth}
\usepackage{textcomp}
\usepackage{xcolor,soul}
\usepackage{color,soul}
\usepackage{algorithm}
\usepackage{algpseudocode}
\usepackage{bm}
\usepackage{esvect}
\usepackage{accents}
\usepackage{caption}
\usepackage{soul}

\begin{document}

\title[MICROSCOPE instrument]{MICROSCOPE instrument description and validation}

\author{ Fran\c{c}oise Liorzou$^1$, Pierre Touboul$^1$,  Manuel Rodrigues$^1$, Gilles M\'etris$^2$, Yves Andr\'e$^3$, Joel Berg\'e$^1$, Damien Boulanger$^1$, Stefanie Bremer$^4$\footnote{Current adress: DLR, Institute of Space Systems, Robert-Hooke-Str. 7, D-28359 Bremen, Germany}, Ratana Chhun$^1$, Bruno Christophe$^1$, Pascale Danto$^3$, Bernard Foulon$^1$, Daniel Hagedorn$^5$, Emilie Hardy$^1$, Phuong-Anh Huynh$^1$, Claus L\"ammerzahl$^4$, Vincent Lebat$^1$, Meike List$^4$\footnote{Current adress: DLR Institute for Satellite Geodesy and Inertial Sensing, Am Fallturm 9, D-28359 Bremen, Germany}, Frank L\"offler$^5$, Benny Rievers$^4$, Alain Robert$^3$, Hanns Selig$^4$\footnote{Current adress: GERADTS GMBH, Kleiner Ort 8, D-28357 Bremen, Germany}}

\address{$^1$ DPHY, ONERA, Universit\'e Paris Saclay, F-92322 Ch\^atillon, France}
\address{$^2$ Universit\'e C\^ote d{'}Azur, Observatoire de la C\^ote d'Azur, CNRS, IRD, G\'eoazur, 250 avenue Albert Einstein, F-06560 Valbonne, France}
\address{$^3$ CNES, 18 avenue E Belin, F-31401 Toulouse, France}
\address{$^4$ ZARM, Center of Applied Space Technology and Microgravity, University of Bremen, Am Fallturm, D-28359 Bremen, Germany}
\address{$^5$ PTB, Physikalisch-Technische Bundesanstalt, Bundesallee 100, 38116 Brunswick, Germany}

\ead{francoise.liorzou@onera.fr, manuel.rodrigues@onera.fr}
\vspace{10pt}
\begin{indented}
\item[] December 2020
\end{indented}

\begin{abstract}
Dedicated accelerometers have been developed for the MICROSCOPE mission taking into account the specific range of acceleration to be measured on board the satellite. Considering one micro-g and even less as the full range of the instrument, leads to a customized concept and a high performance electronics for the sensing and servo-actuations of the accelerometer test-masses. In addition to a very accurate geometrical sensor core, a high performance electronics architecture provides the measurement of the weak electrostatic forces and torques applied to the test-masses. A set of capacitive sensors delivers the position and the attitude of the test-mass with respect to a very steady gold coated cage made in silica. The voltages applied on the electrodes surrounding each test-mass are finely controlled to generate the adequate electrical field and so the electrostatic pressures on the test-mass. This field maintains the test-mass motionless with respect to the instrument structure. Digital control laws are implemented in order to enable instrument operation flexibility and a weak position sensor noise. These electronics provide both the scientific data for MICROSCOPE's test of General Relativity and the data for the satellite drag-free and attitude control system (DFACS).

\end{abstract}


\submitto{\CQG}
%
%
%

\section{Introduction}
The MICROSCOPE (Micro-Satellite \`a tra\^in\'ee Compens\'ee pour l'Observation du Principe d'Equivalence) mission was fully dedicated to the in-orbit test of the Universality of free fall, the so-called Weak Equivalence Principle (WEP) \cite{einstein08, einstein16}. The accuracy objective was expected to be better than $10^{-15}$ \cite{touboul01a}. The test principle consists in comparing the accelerations of two test-masses of different composition in the Earth gravitational field. The payload of the MICROSCOPE satellite is a double electrostatic accelerometer composed of two cylindrical and concentric test-masses made of different materials. The accelerometer measures the difference of accelerations needed to be applied on the two test-bodies to keep their motion identical with respect to each other while undergoing the same gravity. A non-null difference indicates a violation of the WEP. 

Based on a long legacy \cite{touboul99}, MICROSCOPE payload developments re-use some very well-known and mastered processes and technologies concerning the electronics, as well as the accelerometer mechanical cores. The instrument is called T-SAGE which stands for “Twin Space Accelerometer for Gravity Experiment’. The target value at $10^{-15}$ requires space accelerometers with resolution in the femto-g range. This means stringent specifications on both the instrument and its environment, and extensive pre-flight testing to verify that this performance level can be reached once in space. The instrument must be then carefully characterized to identify the effects of any imperfections and to measure its sensitivity to space environment. 

This paper provides a more detailed description of the instrument T-SAGE than the first result papers \cite{touboul17,touboul19}. It shows some steps among the most critical, of verifications, validations and tests performed on ground and when possible, confirmation of the characterized performance during the mission in space.


\section {T-SAGE instrument description}
\subsection {Instrument sub-system and data production} \label{sect:Inst}

The MICROSCOPE payload is actually made up of two independent “Space Accelerometer for Gravity Experiment” (SAGE) sensors. The first one, designed for the WEP test, embarks two test-masses made of different composition, one in platinum rhodium and one in titanium alloys. The other sensor is identical except for the test-masses composition, both in platinum alloy, and is used to check the experiment (from measurement to the data process). Each SAGE is composed of three units described hereafter. 
The Sensor Unit (SU) comprises two concentric cylindrical test-masses each surrounded by a set of electrodes. The sensor with platinum/titanium masses is called SU-EP and the sensor with platinum/platinum masses is called SU-REF. The material composition and main physical parameters of the test-masses are summarised in Tab. \ref{tab:metrology}.

\begin{figure}
\begin{center}
\includegraphics[width=0.85\textwidth]{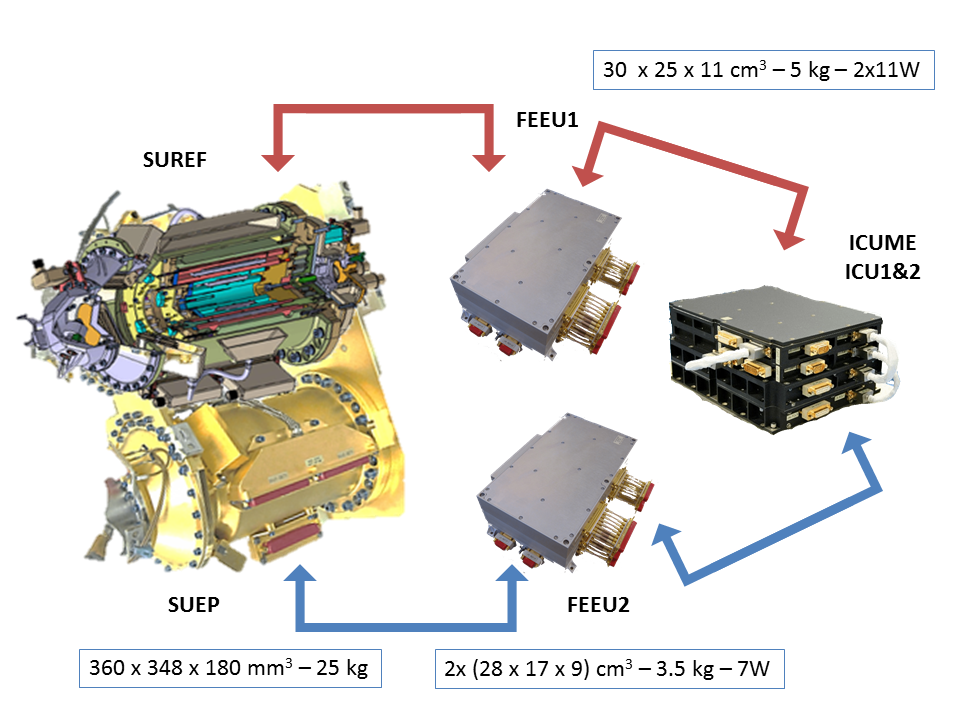}
\caption{View of the T-SAGE elements with their characteristics (volume, mass, consumption).}
\label{fig:tsage}       
\end{center}
\end{figure}

The payload is organised in three units as shown in Fig. \ref{fig:tsage}. The Front End Electronic Unit (FEEU) includes the capacitive sensing of test-mass motion, the reference voltage sources and the analog electronics to generate the electrical voltages applied onto the electrodes. The Interface Control Unit (ICU) hosts the satellite power converters and the digital electronics associated to the FEEU for the servo-control of the test-masses and the interfaces to the data and power bus of the satellite. 

The SU and FEEU have been integrated inside a highly stabilised thermal case in the centre of the satellite while the ICU less thermally demanding is accommodated in one of the satellite wall equipment.

The two-times-six acceleration measurements provided by the accelerometers of one SAGE along and around its three perpendicular sensitive axes are also used in real time on board the satellite by the Drag-Free and Attitude Control System (DFACS)\cite{prieur17}.
The “drag-free point” is defined by a selected combination of the four test-mass acceleration measurements and was regularly changed to the benefit of the experiment validation. The satellite has been in inertial pointing or rotating (called also satellite spin) about the normal axis to the orbital plane during the mission. The angular acceleration measured by the accelerometers was associated to the satellite star-sensor in order to estimate and then to finely control the attitude of the satellite.

The experiment was performed during two years in different configurations in order to assess the instrument sensitivities to the environment. The data delivered by T-SAGE contain scientific measurements for the equivalence principle test (accelerations) and housekeeping measurements to monitor the instrument (electronics modes, reference voltages, temperature, error status, test-mass position and attitude) with precise time. The science data is used in real time by the DFACS. All data are stored on board by the satellite computer before being transmitted to the ground control station. 

\begin{table}
\caption{\label{tab:metrology} Test-masses main physical parameters measured in laboratory before flight. $^{(*)}$ Deduced by density comparison between slices of material extracted at each end of the test-mass during manufacturing.}
\begin{indented}
\item[]\begin{tabular}{@{}lllll}
\br
Parameter & IS1-SUREF & IS2-SUREF & IS1-SUEP & IS2-SUEP \\
\mr
Inner radius [mm] & 30.801 & 60.799 & 30.801 & 60.802 \\
Outer radius [mm] & 39.390 & 69.397 & 39.390 & 69.401 \\
Length [mm] & 43.331 & 79.821 & 43.330 & 79.831 \\
Inertia about $X$ [kg\,mm$^2$] & 125.0206 & 1442.454 & 125.0775 & 319.0266 \\
Inertia about $Y$ [kg\,mm$^2$] & 125.0021 & 1442.139 & 125.0524 & 318.9978 \\
Inertia about $Z$ [kg\,mm$^2$] & 125.0070 & 1442.214 & 125.0549 & 318.9867 \\
Maximum relative difference & 0.0004 & 0.0007 & 0.001 & 0.0001 \\
in moment of inertia \\
Mass [kg] & 0.401533 & 1.359813 & 0.401706 & 0.300939 \\
Density @ 20$^{\rm o}$C [g\,cm$^{-3}$] & 19.967 & 19.980 & 19.972 & 4.420 \\
Density homogeneity along $X^{(*)}$ & 0.04\% & 0.05\% & 0.1\% & 0.001\% \\
\br
\end{tabular}
\end{indented}
\end{table}

\subsection {Operation principle} \label{sect:Ope}
\subsubsection {Servo-loop overview.} \label{sect:loop}
\begin{figure}
\begin{center}
\includegraphics[width=0.8\textwidth]{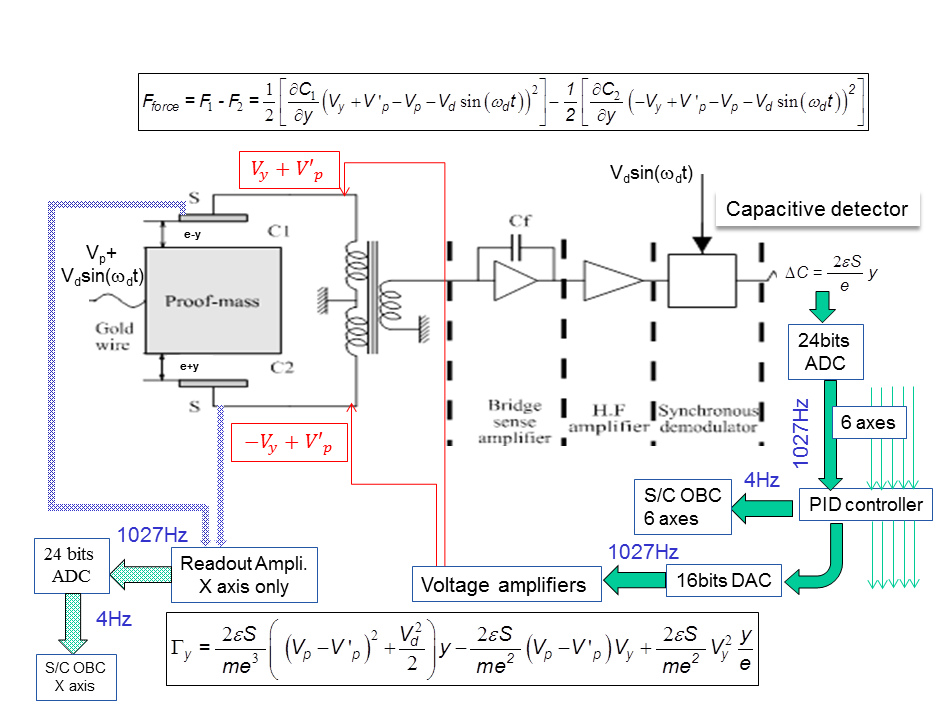}
\caption{Schematic of one degree of freedom servo-loop control.}
\label{fig:loop}       
\end{center}
\end{figure}

The measurement principle of the T-SAGE instrument relies on the levitation of a cylindrical test-mass maintained centred with respect to surrounding electrodes by means of electrostatic forces. The test-mass position with respect to the surrounding electrodes is finely measured by capacitive position sensors. The electrostatic forces are generated from a digital servo-loop controller (Fig. \ref{fig:loop}). The resultant motions of the test-masses arround the Earth follow the same gravitational orbit.

The set of control electrodes is placed on two cylinders inside and outside each test-mass. The configuration of the electrodes is shown in Fig. \ref{fig_xyz} and \ref{fig_Phi}. The translation axes, $Y$ and $Z$,  and the rotation axes $\theta$ and $\psi$ respectively about $Y$ and $Z$, are controlled by the eight electrodes of the inner cylinder, two associated pairs for $Y$ and $\theta$ two associated pairs for $Z$ and $\psi$. The position is obtained by averaging the capacitive measurement of each associated pair while the rotation is calculated by their half difference. The axis of the cylinder, $X$, and the associated rotation, $\phi$, are controlled by the electrodes on the outer cylinder. The measurement of the electrodes $\phi$, four pairs wired together as a single pair, is possible thanks to flat areas on the outside surface of the test-mass which create a variation of the distance between the mass and the electrode when the mass rotates. 

The $X$ axis provides the most sensitive measurements which are used for the equivalence principle test. 

This set of electrodes is used for both measuring the position of the test-mass and applying the control voltage calculated by the servo command laws running at about 1027\,Hz (period of 973$\,\mu$s). This is made possible by the use of a sinusoidal reference voltage $V_d$ (detection) applied on the test-mass as schematised in Fig. \ref{fig:loop}. This is typically a 100\,kHz signal of a few volts; the specific amplitude is chosen in conjunction with the desired detector gain, while the 100\,kHz detection frequency is set much higher than the control loop bandwidth to reduce the sensor noise.

By properly combining the electrodes, six servo-loops channels are controlling independently the translations and attitudes of the test-mass. The operation of the accelerometer is detailed hereafter in section \ref{sect:force}. The signals are digitised in the loop. The selected control law for each mass (each degree of freedom) is a PID (Proportional, Integral, Derivative) configuration. The optimisation of these controllers taking into account the test conditions are made possible by the use of parameters patchable inside the digital electronics of the ICU.

The range of the inertial sensor is limited by the voltage range that can be applied to the electrodes, but must in particular exceed the weak residual accelerations managed by the satellite control. The voltage applied on the electrodes depends on the geometrical and electrical configuration and on the mass of the test-mass (see section \ref{sect:force}).

The value of the test-mass DC voltage, $V_p$, can be set to modify the electrostatic acceleration behaviour: the higher $V_p$, the easier the acquisition of the test-mass but the higher the noise. In order to facilitate the acquisition of the test-mass during the free-fall tests \cite{liorzou14} in ZARM Bremen tower, the $V_p$ voltage and the maximum electrode range voltage have been specifically increased to 90\,V with the Engineering Model Electronics only : component derating rules don't allow this increase on flight models.

Both $V_d$ and $V_p$ voltages are fed to the test-mass through a very thin gold wire described in section \ref{sect_goldwire}.

Scientific measurements are produced by T-SAGE at a frequency of 4Hz, they are of two kinds. The first one is devoted to the satellite Drag-Free and Attitude Control System (DFACS): linear and angular accelerations are deduced from the forces and torques that the controller commands. The second one was originally devoted to the scientific measurement: it is directly picked up on the two $X$ electrodes, in order to reduce the defects of the actuator electronics by the gain of the loop (Readout circuit in Fig. \ref{fig:loop}). The PID corrector is adjusted to fix the electrostatic levitation bandwidth to a few Hz, which is compatible with the measurement frequency needs and with the DFACS  frequency bandwidth. 

\subsubsection {Forces and torques.} \label{sect:force}

\begin{figure} 
\begin{center}
\includegraphics[width=0.45\textwidth]{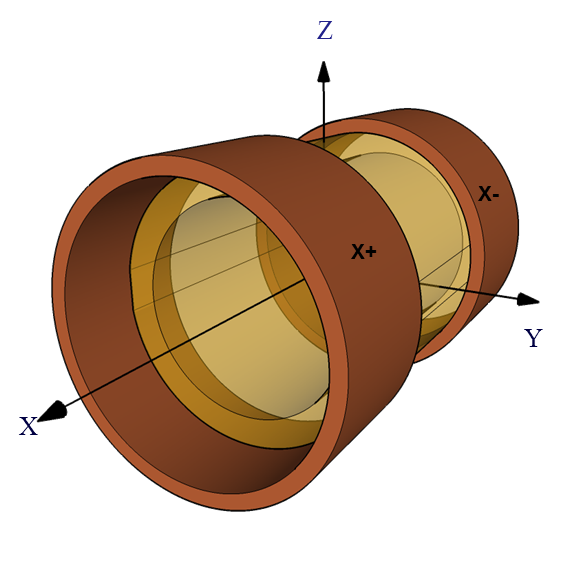}
\includegraphics[width=0.45\textwidth]{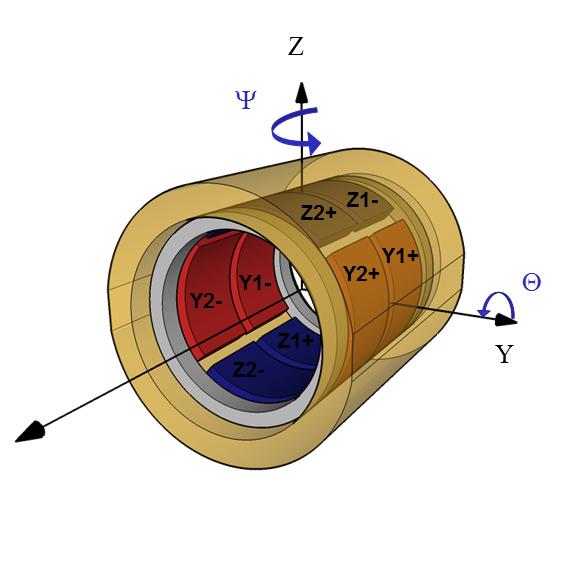}
\caption{Electrode arrangement for $X$ (left), $Y$, $Z$, $\Theta$ and $\Psi$ (right).}
\label{fig_xyz}       
\end{center}
\end{figure}

As shown in Fig. \ref{fig_xyz} the set of $n$ electrodes surrounding the hollow test-mass, inside or outside, exerts a total force along or around any degree of freedom on the test-mass that can be expressed by:
\begin{equation}
F=\sum_n {\frac {1}{2} \nabla (C_{n})(V_n-V_p-v_d)^2}
\label{forcet}
\end{equation}
Where $V_p$ is the DC voltage applied on the test-mass, $V_n$ the voltage applied on the $n$th electrode and $C_{n}$ the capacitance between the $n$th electrode and the test-mass. $v_d=\sqrt2 \, V_d\, \sin(\omega_{\rm 100\,kHz}t)$ is the 100\,kHz voltage applied on the test-mass superimposed to $V_p$: $V_d$ is the root mean value (rms).

For $X$, the expression of the capacitance is established by considering that the electrode is long enough with respect to the front end of the test-mass. A 2D axisymmetric simulation of electrostatic field showed that the field lines are not disturbed anymore when the length of the electrode is greater than 2 times the thickness of the hollow test-mass cylinder. In Fig. \ref{fig_xyz}, there are two electrodes to be considered, $X+$ and $X-$. The capacitance of the two electrodes with respect to the test-mass is given by:
\begin{equation}
    \begin{split}
       C_{x_+}= \frac {2 \pi \epsilon_0 (h+x)} {ln(R_x/R_p)}\\
       C_{x_-}= \frac {2 \pi \epsilon_0 (h-x)} {ln(R_x/R_p)}
    \end{split}
\label{CapaX}
\end{equation}
with $R_x$ the electrode $X$ radius, $R_p$ the test-mass outer radius, $h$ the electrode length covering the test-mass when it is centred. The difference $(C_{x_-}-C_{x_+})$ is representative of the displacement $x$:  $(C_{x_-}-C_{x_+})=-\frac {4 \pi \epsilon_0} {ln(R_x/R_p)}x$. A capacitive sensor transforms the difference of capacitance into a proportional voltage which enters at the input of the digital control loop.

Due to the electrical configuration, the force control along $X$ is obtained by:

\begin{equation}
F_x=\frac {1}{2} [\nabla (C_{{x_+}})(V_{x+}-V_p-v_d)^2+\nabla (C_{{x_-}})(V_{x-}-V_p-v_d)^2]
\label{ForceX}
\end{equation}
When the servo-loop operates, $V_{x+}=v_x+V'_{p_x}$ and $V_{x-}=-v_x+V'_{p_x}$, Eq. \eqref{ForceX} simplifies into:
$F_x=-\frac {4 \pi \epsilon_0}{ln(R_x/R_p)} (V_p-V'_{p_x})v_x$ in the control bandwidth lower than $10\,$Hz,

Indeed, the mean value at low frequency of the AC voltage $v_d$ is null. In this expression $v_x$ is the voltage calculated by the servo-loop to control the test-mass motionless. $V'_{p_x}$ is a DC bias voltage applied to each $X$ electrode. The force divided by the mass $m$ of the test-mass represents the electrostatic acceleration that compensates the other accelerations applied to the test-mass in order to maintain the relative motion null \cite{rodriguescqg1}. The acceleration measurement $\Gamma_x=\frac{1}{m}F_x=\hat{G_{ex}}v_x$ along $X$ is determined by $v_x$ and computed into an acceleration with the definition of the physical gain : $\hat{G_{ex}} = -\frac {4 \pi \epsilon_0}{m \, ln(R_x/R_p)}(V_p-V'_{p_x})$ where $(V_p-V'_{p_x})$ fixes the range of control for the considered geometry.

\begin{figure}
\begin{center}
\includegraphics[width=0.75\textwidth]{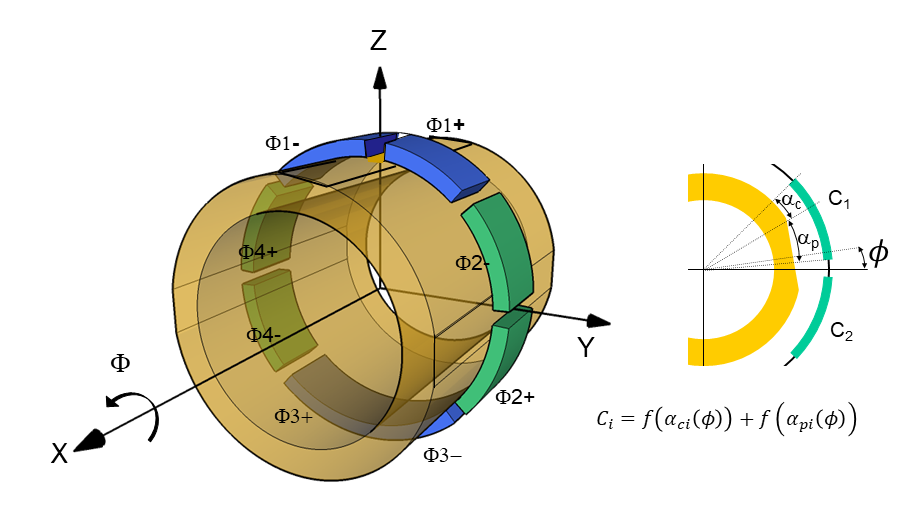}
\caption{Electrode arrangement for $\phi$ measurement.}
\label{fig_Phi}       
\end{center}
\end{figure}

The same principle can be applied to each degree of freedom $w$, the voltage applied to the electrode is: $V_{w\pm{}}=\pm{}v_w+V'_{p_w}$. As shown below, the physical gain becomes proportional to $(V_p-V'_{p_w})$. When the bias $V'_{p_w}$ is null, $V_p$ defines the range of the control for the $w$ axis, otherwise the range of the control is reduced or increased depending on the sign of $V'_{p_w}$. The $V_p$ voltage is a very stable voltage reference applied on the test-mass.  $V'_{p_w}$ is produced by a digital controller and converted by a Digital-to-Analog Converter (DAC) to an analog DC voltage.

For all degrees of freedom, the electrodes are associated by pair: one set in the positive direction and the other one in the negative one. For $Y$ (respectively $Z$), two electrodes $Y1+/-$, $Y2+/-$ (resp. $Z1+/-$, $Z2+/-$) are combined for the position measurement with the help of the capacitive sensing. The capacitance is given by:
\begin{equation}
C_{Y_ks}= \frac {\epsilon_0 S} {e_{\rm in}+s \, y +s (2k-3) \, \psi \, d}
\end{equation}
where $S$ is the area of one $Y$ or $Z$ electrode, $e_{\rm in}$ the gap between inner electrodes and the test-mass, $d$ is the mean distance of the electrode centre from the $Y$ axis (the lever arm of the rotation torque), $k$ denotates the electrode 1 ($k=1$) or 2 ($k=2$) in Fig. \ref{fig_xyz} and $s=+1$ for the electrodes in the positive side of the axis and $s=-1$ for the ones on the negative side.

When the test-mass is controlled by the accelerometer servo-loop to be motionless, $y \ll{}e_{\rm in}$ and $\psi d \ll{}e_{\rm in}$, the difference $\sum_k (C_{Y_{k-}}-C_{Y_{k+}})$ is representative of the displacement $y$ at first order and simplifies to $\sum_k (C_{Y_{k-}}-C_{Y_{k+}}) \approx \frac {4 \epsilon_0 S} {e_{\rm in}^2} y$.

In the same way, the difference $\sum_s {(C_{Y_{1s}}-C_{Y_{2s})}}$ is representative of the angular motion $\psi$ and at first order simplifies to $\sum_s {(C_{Y_{1s}}-C_{Y_{2s})}} \approx \frac {4 \epsilon_0 S \,d} {e_{\rm in}^2} \psi$.

The total force exerted by the four $Y$ electrodes along $Y$ is evaluated with the applied voltage on each electrode ``$ks$'' $V_{yks}=-s\,v_y+s(2k-3) v_{\psi}+ V'_{p_y}$ :
\begin{equation}
F_y \sim -\frac {4 \epsilon_0 S} {e_{\rm in}^2} (V'_{p_y}-V_p)v_y + \frac {4 \epsilon_0 S} {e_{\rm in}^3} ((V'_{p_y}-V_p)^2 + v_y^2+v_{\psi}^2+V_d^2)y
\end{equation}
The linear physical gain along $Y$ of the accelerometer is $\hat{G_{ey}}=-\frac {4 \epsilon_0 S} {m e_{\rm in}^2} (V'_{p_y}-V_p)$. 

The term $\frac {4 \epsilon_0 S} {e_{\rm in}^3} ((V'_{p_y}-V_p)^2 + V_d^2)$ represents the electrostatic stiffness. The term $\frac {4 \epsilon_0 S} {e_{\rm in}^3} (v_y^2)y_0$ represents the quadratic term when the position measurement is biased by $y_0$. Then the servo-loop controlling the output of the PID will apply a force to displace the test-mass to compensate the bias. Finally the term $v_{\psi}^2$ expresses the coupling between the angular control and the linear control.

It has to be noted that at first order with respect to the displacement of the test-mass, there is no stiffness and no non-linear term along $X$.
Actually the back action of the electrodes $Z$, $X$ and $\Phi$ also has an effect on the $Y$ control. The force along $Y$ can be detailed as following:
\begin {equation}
  \begin{aligned}
F_y \sim &-\frac {4 \epsilon_0 S} {e_{\rm in}^2} \frac{\sin(\alpha_y/2)}{(\alpha_y/2)} (V'_{p_y}-V_p)v_y \\
      &+ \frac {2 \epsilon_0 S} {e_{\rm in}^3} (1+\frac{\sin(\alpha_y)}{\alpha_y}) ((V'_{p_y}-V_p)^2 + v_y^2+v_{\psi}^2+V_d^2)y \\
      &+\frac {2 \epsilon_0 S} {e_{\rm in}^3} (1+\frac{\sin(\alpha_y)}{\alpha_y})((V'_{p_z}-V_p)^2 + v_z^2+v_{\theta}^2+V_d^2)y \\
      &+\frac {2 \pi \epsilon_0 R_x h} {e_{\rm out}^3} ((V'_{p_x}-V_p)^2 + v_x^2+V_d^2)y  \\
      &+\frac {\pi \epsilon_0 R_{\phi} L_{\phi}} {e_{\rm out}^3} ((V'_{p_{\phi}}-V_p)^2 + v_{\phi}^2+V_d^2)y
 \end{aligned}
\label {forcey}
\end{equation}
where $e_{\rm out}$ is the gap between the outer electrode cylinder and the test-mass.
If an acceleration is applied along $X$, or a torque applied about $X$, $v_x$ or $v_{\phi}$ is not null and it couples to the measurement along $Y$. This point is discussed in Ref. \cite {chhuncqg5}. The total electrostatic stiffness $K_Y$ depends on the geometry of the inner and outer electrode cylinders :
\begin{equation}
   \begin{aligned}
K_Y &= \frac {4 \epsilon_0 S} {e_{\rm in}^3} ((V'_{p_y}-V_p)^2 +V_d^2) \\
&+\frac {4 \epsilon_0 S} {e_{\rm in}^3} ((V'_{p_z}-V_p)^2 +V_d^2) \\
&+\frac {2 \pi \epsilon_0 R_x h} {e_{\rm out}^3} ((V'_{p_x}-V_p)^2 +V_d^2) \\
&+\frac {\pi \epsilon_0 R_{\phi} L_{\phi}} {e_{\rm out}^3} ((V'_{p_{\phi}}-V_p)^2+V_d^2)
   \end{aligned}
\end{equation}

A similar analysis can be performed on $Z$ axis in which $z$ plays the same role as $y$. We can switch the role of $z$ and $y$ and the role of the associated rotations $\theta$ and $\psi$.

For the angular torque about $Y$ ($\Theta$) or about $Z$ ($\Psi$), the equations are quite similar with $V'_{p_{\psi}}=V'_{p_{y}}$, respectively  $V'_{p_{\theta}}=V'_{p_{z}}$ because $\psi$, respectively $\theta$, is controlled by the same set of electrodes as $Y$, respectively $Z$. The impact of $X$ and $\Phi$ on the rotation has to be considered through the force multiplied by the lever arms defined by the centre of electrodes. For $X$, if a force is exerted by the electrode along $Y$ or $Z$ at a distance $\frac 1 2 (L_p-h)$ (being the distance of the $X$ electrode centre from the $Y$ or $Z$ axis), a torque shall appear. In the same way for $\Phi$ electrodes, the lever arm is $\frac 1 4 L_{\phi}$, where $L_p$ is the test-mass length and $L_{\phi}$ the $\Phi$ electrode length. The total torque for $\Psi$ degree of freedom is thus given by:

\begin {equation}
   \begin{aligned}
T_{\psi} \sim &-\frac {4 \epsilon_0 S \, d} {e_{\rm in}^2} (V'_{p_y}-V_p)v_{\psi} \\
                    &+ \frac {4 \epsilon_0 S\, d} {e_{\rm in}^3} \left((V'_{p_y}-V_p)^2 + v_{\psi}^2+v_y^2+V_d^2\right)\psi \\
                    &+\frac {4 \epsilon_0 S\, d} {e_{\rm in}^3} \left((V'_{p_z}-V_p)^2 + v_z^2+v_{\theta}^2+V_d^2\right)\psi \\
                    &+\frac {2 \pi \epsilon_0 R_x h} {e_{\rm out}^3}\frac {(L_p-h)}{2} \left((V'_{p_x}-V_p)^2 + v_x^2+V_d^2\right)\psi \\
                    &+\frac { \pi \epsilon_0 R_x L_{\phi}} {e_{\rm out}^3} \frac {L_{\phi}} {4}  \left((V'_{p_{\phi}}-V_p)^2 + v_{\phi}^2+V_d^2\right)\psi
   \end{aligned}
\label{torquey}
\end{equation}

The physical gain of the angular acceleration $T_{\psi}/J_z$ is given by : 
$\hat{G_{e\psi}}=-\frac {4 \epsilon_0 S \, d} {J_z e_{\rm in}^2} (V'_{p_y}-V_p)$ 
where $J_z$ is the mechanical inertia moment of the test-mass about $Z$. \\

For the control in rotation about $\Phi$, the angular motion is defined by the sum of the difference of capacitance of the 4 pairs $\Phi$ electrodes (Fig. \ref{fig_Phi}) evaluated at first order to : $\Delta C_{\phi} \approx 16 \epsilon_0 \sin(\frac {\alpha_e}{2}) R_x \frac {L_{\phi}}{e_{\rm out}^2}\phi$. $\alpha_e$ is the angle defining the arc of the electrode in Fig. \ref{fig_Phi} ($\sim 2\pi/8$ if we neglect the spaces between the electrodes). The torque is expressed by: 
\begin{equation}
T_{\phi} \sim \frac {8 \epsilon_0 L_{\phi} \, R_x} {e_{\rm out}} \left[1-\frac {1+\cos(\alpha_f /2)}{1-\cos(\alpha_f /2)}\right](V'_{p_{\phi}}-V_p)v_{\phi}
\end{equation}
where $\alpha_f$ is the angle defining the flat area on the test-mass. In this particular case of $\Phi$ degree of freedom, the physical gain converting the applied voltage to an angular acceleration becomes: $\hat{G_{\phi}}=\frac {8 \epsilon_0 L_{\phi} \, R_x} {J_x e_{\rm out}} \left[1-\frac {1+\cos(\alpha_f /2)}{1-\cos(\alpha_f /2)}\right](V'_{p_{\phi}}-V_p)$. The test-mass has been defined with a particular relationship between its length and its radiuses in order to have all moment of inertia identical about all axes $J_x \sim J_y \sim J_z$. This equality is questioned by the accuracy of the machined parts and the homogeneity of the material given in Tab. \ref{tab:metrology}.
In the following, the scale factor is defined as the ratio of the estimated physical gain $\hat{G_{ew}}$ to the real one $G_{ew}$, and should be close to 1: $K_{1w}=\frac {\hat{G}_{ew}} {G_{ew}}$ ($w=x,y,z,\phi, \theta$ or $\psi$).

\begin{table}
\caption{\label{tab_gain} Theoretical physical gain and position sensitivity for each degree of freedom used in flight for the DFACS and for the ground segment.}
\begin{indented}
\item[]\begin{tabular}{@{}lcccccc}
\br
Physical gain & $X$ & $Y$ & $Z$ & $\phi$ & $\theta$ & $\psi$  \cr
&\multicolumn{3}{c}{$\mu$m\,s$^{-2}$/V} &\multicolumn{3}{c}{$\mu$rad\,s$^{-2}$/V} \cr
\mr
IS1 SU-REF & 0.0688 & 0.161  & 0.161 & 2.09 & 4.65 & 4.65 \cr
IS2 SU-REF & 0.0538 & 0.438 & 0.438 & 1.29 & 7.03 & 7.03 \cr
IS1 SU-EP   & 0.0689 & 0.161 & 0.161 & 2.09 & 4.65 & 4.65 \cr
IS2 SU-EP   & 0.0805 & 0.992 & 0.991 & 5.79 & 1.57 & 1.58 \cr
\br
Position sensor gain & $X$ & $Y$ & $Z$ & $\phi$ & $\theta$ & $\psi$  \cr
&\multicolumn{3}{c}{$\mu$m/V} &\multicolumn{3}{c}{$\mu$rad/V} \cr
\mr
IS1 SU-REF & 3.33 & 4.35 & 4.35 & 708 & 493 & 493 \cr
IS2 SU-REF & 3.85 & 3.23 & 3.23 & 100 & 191 & 191  \cr
IS1 SU-EP   & 3.33 & 4.35 & 4.35 & 708 & 493 & 493 \cr
IS2 SU-EP   & 3.85 & 3.23 & 3.23 & 100 & 191 & 191 \cr
\br
\end{tabular}
\end{indented}
\end{table}

\subsection {Sensor Unit (SU) descriptions} \label{sect:Sensor}
The two Sensor Units SU-REF and SU-EP are fixed on a mechanical interface with the satellite, named SUMI (Fig. \ref{fig_core}). In order to allow the sensor alignments during the integration on the satellite, an optical cube is implemented on the SUMI as well.

\begin{figure}
\begin{center}
\includegraphics[width=0.95\textwidth]{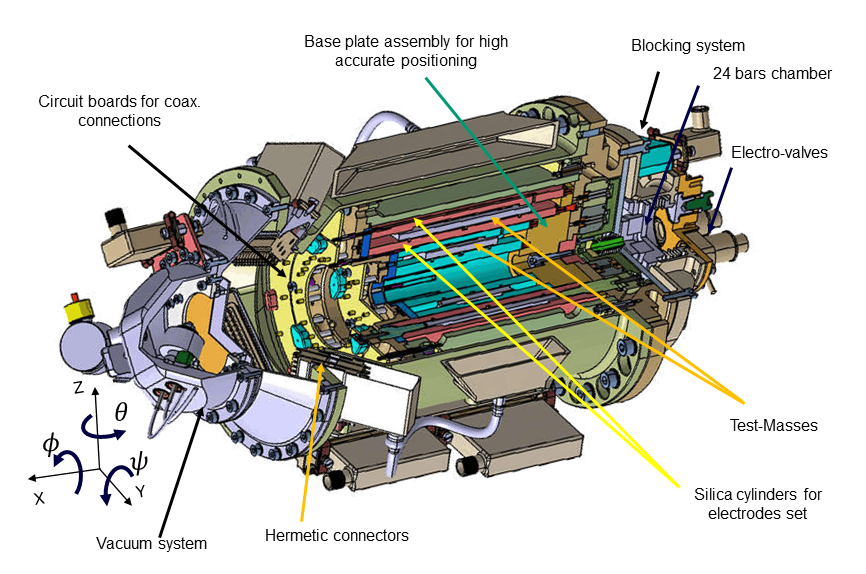}
\caption{Cross-cut view of the differential accelerometer sensor unit, comprising two concentric accelerometers in a tight housing with the blocking mechanism on the bottom of the SU and the vacuum system at the top of the SU.}
\label{fig_su}       
\end{center}
\end{figure}

Each Sensor Unit is a mechanical assembly of the following elements (Fig. \ref{fig_su}):
\begin{itemize}
\item Two test-masses: they serve as inertial reference for the drag-free and attitude control system and as mass probes for the test of the equivalence principle;
\item The electrode set for the capacitive sensing and for the electrostatic control of the test-masses; they are made of gold coated silica cylinders;
\item A mounting base-plate made of gold coated silica for a precise positioning of the electrode sets;
\item A sole plate made of Invar for the positioning of the mounting base plate, and which supports the titanium inner housing and the vacuum tight housing;
\item Mechanical stops that prevent the test-mass to be electrically in contact with the electrodes and that take the shock of any mechanical contact to preserve the silica electrode set; 
\item One 7$\mu$m gold wire connecting each test-mass used to apply a very stable voltage on the test-masses;
\item A test-mass blocking mechanism in order to prevent any damage from the heavy test-masses (0.4\,kg to 1.4\,kg) because of launch phase vibrations and micro-satellite in-orbit release shocks. The test-masses are unlocked before the first in-orbit switch on;
\item A vacuum system to maintain a sufficiently low vacuum and cleanliness of the electrode set and to ensure the performance of the instrument;
\item A vacuum Invar tight housing encloser that closes hermetically the SU;
\item Hermetic electrical connectors.
\end{itemize}

\subsubsection {Test-masses.}

A particular effort has been paid on the definition of the test-mass shape. This shape guarantees a certain low level of sensitivity to gravity disturbances in the acceleration differential measurement. 
The gravity acceleration module $g^{(i)}$ undergone by the $i$th test-mass of mass $m^{(i)}$, considered perfect, from a source mass $M_s$ at distance $R_s$ is expressed at order 2 of the mechanical moments in \cite{willemenot97}:
\begin{equation}
g^{(i)} \approx \frac {G M_s}{R_s^2} \left[1+\frac {3}{2 m^{(i)} R_s^2}(J_y^{(i)}+J_z^{(i)}-2 J_x^{(i)})+...\right]
\end{equation}
where $J_x^{(i)}$ , $J_y^{(i)}$ and $J_z^{(i)}$ are the mechanical moment of inertia around the respective test-mass axes $X$, $Y$ and $Z$. When measuring the differential acceleration between two perfect concentric test-masses, the term $g^{(1)}-g^{(2)}$ does not totally disappear \cite{rodriguescqg1}. A term depending on the geometry remains at first of order:
\begin{equation}
g^{(1)}-g^{(2)} \approx \frac {G M_s}{R_s^2} \frac {3}{2R_s^2} \left[\frac{1}{m^{(1)}} \left(J_y^{(1)}+J_z^{(1)}-2 J_x^{(1)} \right) - \frac{1}{m^{(2)}} \left(J_y^{(2)}+J_z^{(2)}-2 J_x^{(2)}-2 J_x^{(1)}\right)\right]
\label{eq_grav}
\end{equation}

Equation \eqref {eq_grav} shows that for a sphere, with all mechanical moments being identical, the gravity field should not depend any more on the test-mass geometry. For cylindrical test-masses, by adjusting the length for a giving radius to get $\left(J_y^{(i)}+J_z^{(i)}-2 J_x^{(i)}\right)=0$, it is possible to cancel the effect of local gravity at first order.  

\subsubsection {Sensor core: electrode cylinders and sole plate.}
\begin{figure}
\begin{center}
\includegraphics[width=0.3\textwidth]{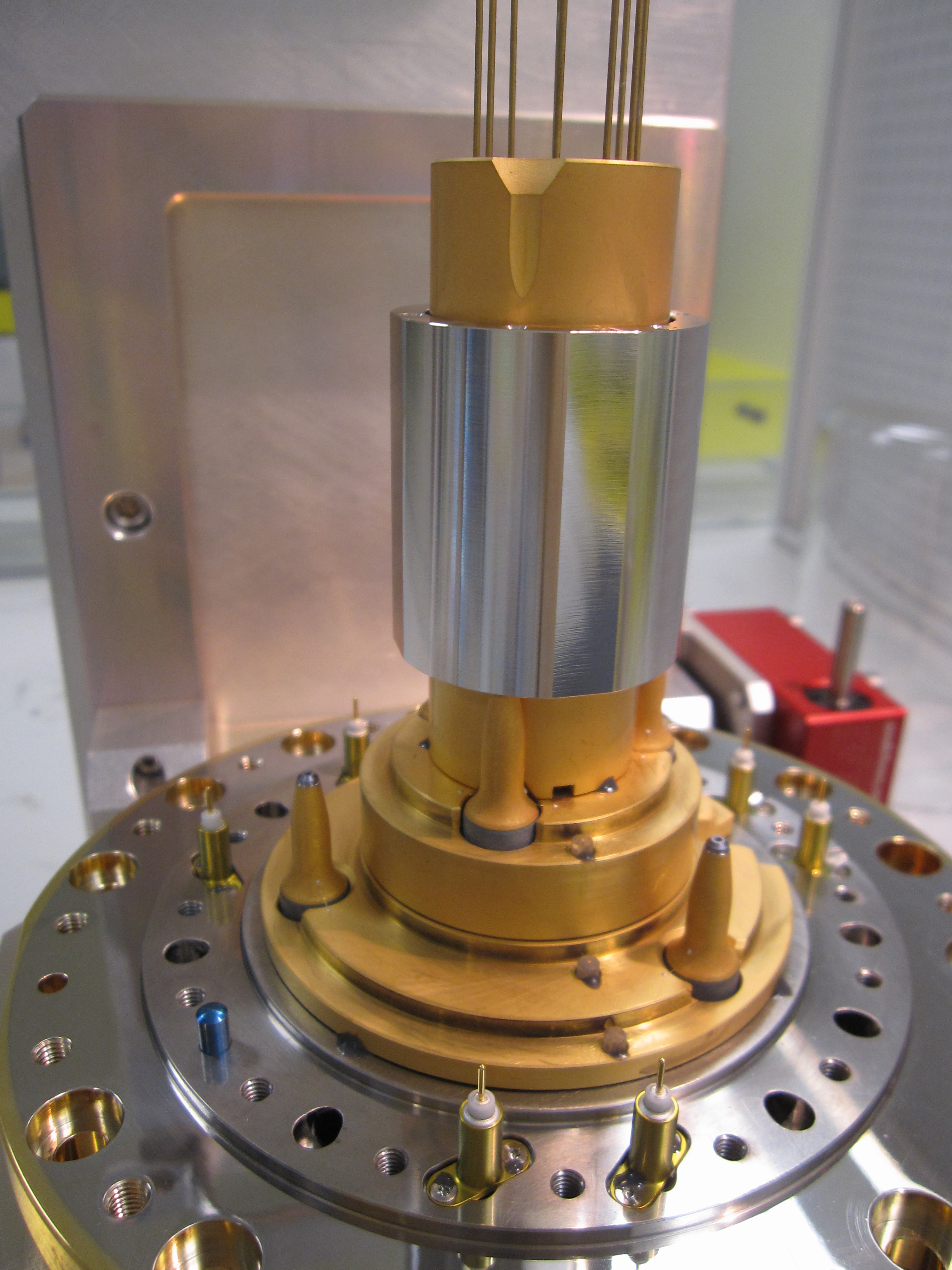}
\includegraphics[width=0.3\textwidth]{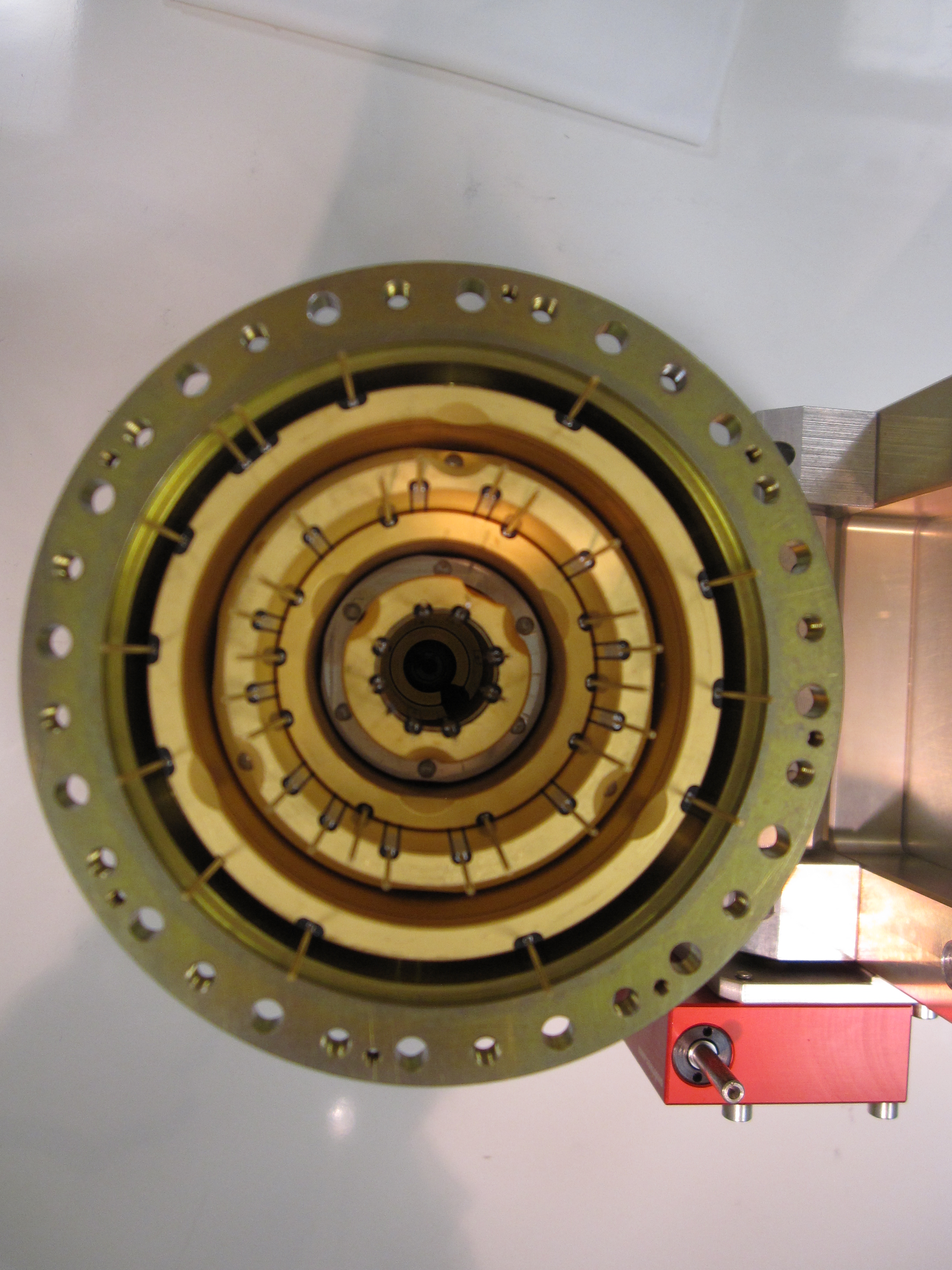}
\includegraphics[width=0.35\textwidth]{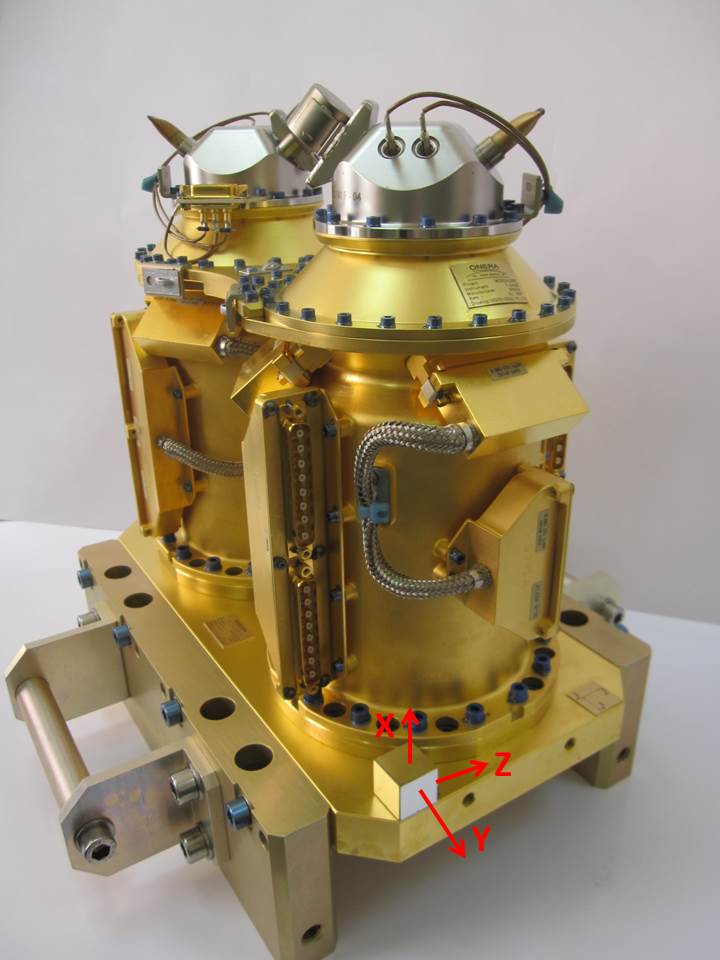}
\caption{Left: step of integration of the inner SUREF test-mass laying on the stops and surrounding the first silica gold coated cylinder laying on the silica reference base-plate. This base-plate is mounted on the polished Invar sole plate. Middle: all gold coated cylinders integrated around the two test-masses of the SUEP. Right: SUEP and SUREF fixed on the SUMI titanium gold coated plate. SUREF is the SU closer to to the optical cube. Coaxial shielded connectors deliver all the capacitive sensing information.}
\label{fig_core}       
\end{center}
\end{figure}

Each test-mass is surrounded by two cylinders (one inner, one outer) made of silica (Suprasil 2\textregistered{}). The cylinders are engraved by ultrasonic machining (ONERA patent) and gold coated in order to provide the necessary electrodes to control the six degrees of freedom of each test-mass. They are integrated on the Silica base-plate, also accurately machined and gold coated, see Fig. \ref{fig_core}. The ultrasonic machining ensures an accuracy up to $5\,\mu$m. The control of the silica parts is performed  at $1\,\mu$m accuracy with a 3D metrology system.
All silica parts are accurately mounted on an Invar\textregistered{}\,sole plate which serves as reference for all the core centring and alignment. The sensor core is closed by an inner housing and a clamp. The sole plate also hosts the mobile parts of the blocking system.

Internal circuit boards complete the mechanical core of the double accelerometer. 
This core is closed by a hermetic tight housing composed of the outer housing, the hermetic connector support and the vacuum management system. 

The vacuum is ensured passively in flight thanks to a getter material ST172\textregistered{}. Its porous structure allows the gases to diffuse inside the getter where a sorption mechanism acts irreversibly for the molecules of CO, CO$_2$, O$_2$, N$_2$ and in a reversible way for those of H$_2$. An ionic pump is used on ground until delivery. Through a resistive wire embedded in the material, the material is heated to a temperature of 900\,\textdegree{}C during 10 minutes. This is the activation process that is applied when integrating the SU. Then during the life of the instrument, the surface of the getter is exposed to a gradual saturation due firstly to the residual degassing of the internal parts to the sealed enclosure (first months after closing) and secondly to the porosity of the materials of the enclosure and especially ceramics of the various sealed crossings (storage on the ground until launch). Even if the leakage rate level of the hermetic housing and connectors is very low (better than $10^{-10}$\,mbar\,l\,s$^{-1}$) and ensures the low vacuum level inside the sensor unit, the getter has been periodically reactivated by heating to restart the process of diffusion into the material and clean the surface.

In orbit, the vacuum has been specified to be lower than $10^{-5}$\,Pa and the outgassing rate inside the housing to be lower than $3\times{}10^{-10}$\,Pa\,m$^3$s$^{-1}$. A test, performed on a representative model from January 2007 to September 2015, showed that the pressure was maintained under the specification during 600 days over 930 days of test, only limited by the time between two reactivations of the getter. This test duration largely covers the ground test and pre-launch storage period.
The hermiticity of the mechanical interfaces is ensured by indium seals laid on the polished surfaces.

The external housing supports the electrical connection of the electrodes, the test-masses and the temperature probes. This housing has been tested with $10^{-9}$\,mbar\,l\,s$^{-1}$ ($10^{-13}$\,Pa\,m$^{-3}$s$^{-1}$) tightness. The connectors are D-subminiature to enable the transmission of more than 40 shielded signals assembled in a specially developed circuit board.

\subsubsection {Gold wire.} \label{sect_goldwire}
A 7$\mu$m diameter gold wire is connected to each test-mass. The goal of such a mechanical link is to control the test-mass charge in order to avoid any parasitic forces fluctuation due to the flux of charged particles in orbit and to keep as stable as possible the voltage on the test-mass and thus the scale factor. The gold wire is thin and as long as possible to minimise the resulting stiffness and damping. The wires are made of gold (rust free) with the main characteristics in Tab. \ref{tab_wire}.

\begin{table}
\caption{\label{tab_wire} Gold wire characteristics.}
\begin{indented}
\item[]\begin{tabular}{@{}ll}
\br
Material & pure gold stabilised with 5\,ppm of beryllium \cr
\mr
Diameter & $7\,\mu$m \cr
\mr
Length & $\sim20$\,mm \cr
\mr
Elongation & $0.5$\% \cr
\mr
Tensile strength & $1$\,cN \cr
\mr
Electrical conductivity & $870\,\Omega$\,m$^{-1}$ \cr
\br
\end{tabular}
\end{indented}
\end{table}

This wire is implemented along the sensitive $X$-axis: the first end is glued with electrically conductive adhesive on the bottom of the test mass, and the other end is glued on a pin fixed on the base plate for the outer test-mass and on a specific electrode for the inner one.

In orbit, before the release of the test-masses, the conduction between the masses and the sources of references $V_p$ and $V_d$ is done both through the gold wire and through contact with the mechanical stops. In order to verify the integrity of the gold wire and thus their resistance to the launch vibrations, an electrical test is performed. A remote control command opens the connection to the stops through a relay. In presence of the gold wire, the capacitive sensing keeps on giving the same signal representative of the test-mass position since $V_d$ is still coming through the gold wire. In case of gold wire failure, when the relay of the stops is open, $V_d$ is not applied on the test-mass anymore neither through the gold wire nor through the stops: the detector signal drops to zero. This simple test was successfully performed in orbit at the beginning of the mission to green light moving towards the next step, the commissioning phase.

\subsubsection {Blocking system and mechanical stops.}
The blocking system limits the test-mass motion during the launch vibrations.  It is retracted once in orbit in order to release the test-masses and start the instrument operation. As shown in Fig. \ref{fig_su}, it consists of three fixed stops at the top end of each mass and three mobile stops at the bottom end of each mass. The rounded end of each stop fits into a corresponding conical hole in the test-mass. The shapes of both stop and hole are designed to prevent the mass from sticking to each other after a prolonged contact under the locking force. The retractation system consists of a pressure vessel which applies a total $2500$\,N locking force on the stops. Once in orbit, an electro-valve is activated to open the vessel to the space vacuum, which allows a compressed spring to push away the mobile stops. They are retracted by $150\,\mu$m, which allows for a theoretical free motion of the test-mass of $\pm{}75\,\mu$m. As the stops are not fully retracted from the holes the radial motion is also limited to the same $\pm{}75\,\mu$m value thanks to the conic shape of the holes. The limited free motion ensures an easier electrostatic levitation of the test-mass in orbit. At the same time, the gap between the stops and the mass must be as large as possible to limit the patch effect disturbances.

\subsection {Electronics description} \label{sect:Elec}
The payload electronics puts up together different functional sub-systems distributed over the FEEU or the ICU boxes.

The Interface Control Unit (ICU) contains:
\begin {itemize}
\item The Power conversion units (PCU) delivering 6 secondary voltage buses (5\,V, 3.3\,V, $\pm{}$15\,V, $\pm{}$48\,V) with line regulation and low noise and temperature sensitivity. The PCU function is the only one to have a cold redundancy;
\item The Digital Signal Processing (DSP) board for both the computation of the servo-control signals and the interface with the satellite;
\item A serial high-speed (RS422 bi-directional) link that transmits back and forth between FEEU and ICU the position data and actuation control signals at a baud-rate of 1.25\,Mbps;
\item The telemetry and command control with the satellite.
\end {itemize}

The Front End Electronics Unit (FEEU) comprises:
\begin {itemize}
\item The Front-end capacitive position sensors and the amplifiers for the electrostatic actuation;
\item The voltage references ($V_p$, $V_d$) for the test-masses, and the ones for the Analog to Digital Converters (ADC).
\end{itemize}

The electrode signals, the test-mass voltages and the temperature probe signals are carried out between the SU and the FEEU through three harnesses of more than 40 pico-coaxial cables. The harnesses exhibit low thermal conductivity and stray capacitance compatible with the tuning of the capacitive bridge and the space outgassing requirements. 

\subsubsection {Position detector: Capacitive sensing.}
Capacitive detectors are contact-less, high-resolution position sensors \cite{josselin99}.  They consist in an electromechanical transducer and a conditioning electronic circuit. The electromechanical transducer is constituted by a couple of electrodes facing the test-mass opposite surfaces. The displacements of the test-mass modulate the voltage difference between the electrodes and the test-mass set to a $100$\,kHz sine voltage $V_d$. The displacements induce thus a variation of charge that is sensed by the detector conditioning electronics. The conditioning electronics is based on a 100\,kHz synchronous detection scheme whose pumping frequency is far beyond the displacement bandwidth. Capacitive detectors can thus be implemented on the same set of electrodes as that used for actuation. The first stage of the electronics is a differential transformer linked to the sensor electrodes. The differential transformer and the capacitances of the test-mass/electrodes pair form a capacitive bridge. It is terminated by the feedback capacitance of a charge preamplifier. The preamplifier stage carries out a charge to voltage transformation resulting in a scale factor from 16 to 80\,V/pF after amplification by one or two band-pass filters. The detector front-end is followed by a synchronous detection stage (demodulation of the 100\,kHz signal) and a DC amplifier that implements anti-aliasing filtering. Finally, the output stage of the capacitive position sensor is based on an analog-to-digital converter (24-bit $\Sigma \Delta$ ADCs) which delivers position data to the control laws.

\subsubsection {Actuation voltage.}
The drive voltage amplifiers deliver two opposite voltages to opposite electrodes through a superimposition network and from the outputs of the control laws, delivered by a DAC. These amplifiers are specified to be very symmetric and are mainly composed of a first stage providing the open gain of the amplifier and a push-pull circuit providing the level of the necessary high voltage ($\pm{}40$\,V).

\subsubsection {$V_p$ reference voltage.}
The role of the $V_p$ voltage is to linearise the electrostatic actuation as seen in section \ref{sect:force}.
In High-Resolution Mode (HRM), the $V_p$ voltage is set to 5\,V to minimise the back-action of the position sensor noise through the electrostatic stiffness phenomenon while in Full-Range Mode (FRM), the $V_p$ voltage is set to 42\,V to increase the accelerometer full-scale range. In FRM, all $V_p'$ are set to 0\,V. 
In addition to that, the $V_p$ voltage can be set to 7.5\,V for characterisation purposes.

\subsubsection {$V_d$ detection voltage.}
The $V_d$ generator is an RC-oscillator at 100\,kHz whose amplitude is servo-controlled. The amplitude is set to 5\,V$_{\rm rms}$ in HRM measurement phase and to 1\,V$_{\rm rms}$ in all other modes in particular in FRM for the test-mass acquisition. In FRM, the gain of the detectors is reduced to increase the range of the position sensor.

\subsubsection {Analog-to-digital conversion.}
The $2\times{}6$ position signals delivered by the capacitive detectors are digitised before being fed to the servo-control.
Twelve analog-to-digital converters (24-bit $\Sigma\Delta$ ADC7712) simultaneously sample the position signals to ensure synchronisation of the measurements along the 6 degrees of freedom. The sampling frequency, 1027\,Hz, is consistent with the control bandwidth.
As a $\Sigma\Delta$ converter, the AD7712 relies upon a master clock at 10\,MHz that is distributed to all the FEEU ADCs. Each chip sets its Data Ready output to low when the acquisition is completed, i.e. simultaneously after a fixed number of master clock periods. A Field Programmable Gate Arrays (FPGA) polls all the ADCs simultaneously, deserialises the data and forms the packet to be transmitted to the ICU.
For the digital to analog conversion, the MAX542 (Serial link, 16-bit quantisation, “latch” input for synchronisation, low power) is used.
At reception of the ICU packet, the FEEU FPGA dispatches the data to the 24 DACs, the conversion is latched until the ADCs have completed their acquisition. The delay between the acquisition of position and the actuation is therefore deterministic without skew. The total delay, comprising the FEEU-ICU transmission, the ICU processing, the ICU-FEEU transmission equals 2 sampling periods (2/1027\,sec).

\subsubsection {ICU DSP board}
The architecture of the DSP board is based on the radiation-hardened Atmel\textregistered{}\, version of the SHARC ADSP 21020, the TSC21020F model. The DSP board is mounted with the typical peripherals PROM, EEPROM, and SRAM. The input-output constraints of the DSP (separate bus for program and data memory) made it necessary to implement a companion FPGA which in addition to the handling of the memory, takes in charge the interfaces with the satellite computer and the FEEU. The FPGA is an RT54SX72-S from Actel\textregistered{}, hardened component insensitive to latch-ups and upsets (triplication voted internally on an anti-fuse base). The DSP has a 20\,MHz cycle and operates the servo loop at a submultiple of this frequency: 1027\,Hz. 

\subsubsection {ICU Software}
The DSP software of the payload is split into two pieces of code:
\begin{itemize}
\item The Initialisation Software, which is stored and executed in PROM. It enables the interface with the on board computer, loads the Application software and if ordered by a specific telecommand (TC) completes the patch of the Application software in EEPROM.
\item The Application software stored in EEPROM and executed in SRAM. It handles the digital tasks of the accelerometers in a deterministic and cyclic way as: 
     \begin{itemize}
     \item receiving data from position sensors;
     \item applying the control laws;
     \item transmitting the actuation signals;
     \item processing telemetries;
     \item handling asynchronous or synchronous TC that come from the satellite on-board computer (OBC). 
     \end{itemize}
\end{itemize}

\subsubsection {On-board computer interface and telemetry package.}
The data produced by the instrument are transmitted and synchronised to the on-board computer by a serial link. They are transmitted in two telemetry packets at 1\,Hz for the “housekeeping data” and at 4\,Hz for the science data.

The housekeeping or maintenance data are processed to track the good operation of the instrument. Integrity checks are performed and reported to ensure the consistency of the transmitted TM data. They contain the position and attitude of the test-masses, the $V_p$ and $V_d$ voltage monitoring, the SU, the FEEU and ICU temperatures, the synchronisation data, the FEEU and ICU status (configuration, mode, flags),  the secondary supply voltage monitoring, the SU force sensor used before test-mass release, the Single Event Upset (SEU) counter detected in SRAM and the Overload counter.\\

The 4\,Hz sampled science telemetry, called  $TM_{\rm 4Hz}$, contains the acceleration measurement data:
\begin{itemize}
\item 12 degrees of freedom data are taken at the output of the PID controls (at 1027\,Hz), filtered through a low pass infinite impulse response filter and averaged through a simple averaging filter. A priori scale factor is applied to generate an acceleration data directly usable on board by the DFACS;
\item 2 specific out-of-the-loop measurements for the $X$ axis. An additional ADC per test-mass measures the actual voltage applied to the $X$ electrodes. This ADC operates synchronously with the PID control output ones. Therefore, the processing chain is executed in parallel with those of the other axes. The signal path is basically the same; however, the requirements for anti-aliasing are different. There is a separate processing chain for this out of the loop measurement which is transmitted in $TM_{\rm 4Hz}$ packets after the PID output data.
\end{itemize}
The $TM_{\rm 4Hz}$ packets are formed four times per second with a phasing parameterised with respect to the 1\,Hz clock signal delivered to the ICU by the On board Computer. The real-time cycle of the ICU is set by the ADCs the FEEU. Moreover as the integration time of ADCs $T_{\rm ADC}=512\times{}19 /f_{clock}\approx 973\mu s$, with $f_{clock}=10\,$MHz the ADC frequency clock signal, the theoretical number of samples in a $TM_{\rm 4Hz}$ packet is close to 257. Indeed, the number of samples is an integer closest to the ratio $250ms/T_{\rm ADC}=256.99013$. The number of samples is thus 257 most of the time. But as the ratio of clocks is not an integer, this means that the 257$^th$ sample of the FEEU arrives for each $TM_{\rm 4Hz}$ packet closer and closer to when the $TM_{\rm 4Hz}$ packet is formed. There comes a time when this 257$^th$ sample arrives "too late" to be integrated. This is when the ICU "purges" the delay by considering only 256 samples for its packet. The delay is thus reset to zero and the next packet has 257 samples. With perfect setting of the clocks the occurrence of 256 samples in a $TM_{\rm 4Hz}$ is 1 over 102 $TM_{\rm 4Hz}$. In flight due to the FEEU and ICU clock tuning, the occurrence is one over 150 for the SU EP and 1 over 128 for the SU REF. The specification was set to an occurrence greater lower than one over 100, and the measurements show some margins with the needs.


\section{T-SAGE instrument sub-systems test and performance assessment} \label{sect_tsage}

\subsection{SU on-ground verification of T-SAGE instrument } \label{sect:SUground}

\subsubsection {Test mass metrology.}
The cylindrical test-masses have been finely produced and characterised in the laboratory of Physikalisch-Technische-Bundesanstalt (PTB), in Brunswick, with accuracy better than $2\,\mu$m \cite{hagedorn13}. The masses have been measured with a maximum error of 0.025\,mg. The density was evaluated by cutting two slices at the ends of the real test-masses (giving also an estimation of the homogeneity of the obtained material) and measuring the density of the slices at accuracy better than 0.001\,g\,cm$^{-3}$.

The test-masses have 4 small flat areas along $X$ to break the cylindrical symmetry and to manage the angular control of the test-mass about $X$ ($\Phi$ axis) while the length of the test-mass is optimised to keep an inertia momentum identical about the 3 axes. The inertia momentums have been calculated taking into account the measured dimensions and densities. The dispersion of the measured geometrical characteristics, in addition to the measured density, leads to a total relative dispersion of the inertia with respect to the ideal ‘test-mass’ behaving like a homogeneous sphere: the worst case of relative deviation has been evaluated to $10^{-3}$ and is sufficient to minimise the effect of local gravity gradients \cite{rodriguescqg1}. A synthesis of the test-mass metrology data is summarised in Tab. \ref{tab:metrology}, in good agreement with the geometry specifications up to $2\,\mu$m tolerance deviations.

\subsubsection {Core metrology: test-mass centring.}
\label{centering}
The centring of the two concentric test-masses once levited in orbit is specified to $20\,\mu$m \cite{rodriguescqg1}. This centring is achieved by the accuracy of realisation of the different parts and their integration accuracy that leads to produce a symmetric electrostatic field of one test-mass with respect to the other. The integration of all parts has been realised with several critical steps of 3D machine control of the partially assembled parts with up to $1\,\mu$m accuracy. 
The capacitive measurement during the integration process completes the information on the geometry metrology.

By positioning the $X$ axis vertically up or down under the Earth's gravity field, the test-masses rest on the stops at a distance of $\pm{}75\,\mu$m for the SUEP and at $+90/-75\,\mu$m for the SUREF with respect to the theoretical geometric centre. The capacitances of $X$ electrodes are estimated by $C_{+}(x)= \frac{2\pi\epsilon_0}{ln(R_x/R_p)}x+C_{0_{+}}$  and $C_{-}(x)= -\frac{2\pi\epsilon_0}{ln(R_x/R_p)}x+C_{0_{-}}$ and compared to actual measurements for different test-mass displacements $x$ under gravity. $C_{0_{+}}$ and $C_{0_{-}}$ are the capacitance values when the test-mass is placed at the geometrical center ($x=0$). It comprises also border effects mainly due to chamfers which bias the capacitance centre value. The measurement of the capacitances in the two positions associated to the knowledge of the geometry gives the position where the capacitances should be balanced between $C_{+}(x)$ and $C_{-}(x)$: this is at this position that the servo-loop controls the test-mass once in orbit. The balance is not the geometrical centre but a position deduced by the capacitance measurements in different positions. For each test-mass, the position $x_0$ verifying  $C_{+}(x_0) = C_{-}(x_0)$ is given by:
\begin {equation}
\label{eqXep}
x_0=\frac{1}{2\alpha} [\frac{1}{2} (C_{-} (d_1)+ C_{-} (d_2))-\frac{1}{2} (C_{+} (d_1)+ C_{+} (d_2)]-(d2+d1)
\end{equation}
where $\alpha=\frac{2\pi \epsilon_0}{ln(R_x/R_p)}$, $d_1$ is the actual position corresponding to the theoretical displacement of $+75\,\mu$m for the SUEP and $+90\,\mu$m for the SUREF. In the same way,  $d_2$ is the actual position corresponding to the theoretical displacement of $-75\,\mu$m for the SUEP and the SUREF. Because of the accuracy of the mechanical parts (the stops in particular) and of the integration of the stops, $d_1$ and $d_2$ are deduced from the metrology with a few micrometers accuracy. Eq. (\ref{eqXep}) assumes that the equation of perfect concentric infinite cylinders applies. The term $\alpha$ can also be deduced from the measurements:
\begin {equation}
\label{eqXref}
\alpha=\frac{1}{2(d_1-d_2)} [(C_{+} (d_1)- C_{+} (d_2))-(C_{-} (d_1)- C_{-} (d_2)]
\end{equation}
When replacing Eq. (\ref{eqXref}) in Eq. (\ref{eqXep}), the offcentering is deduced without strong assumption on the capacitance formulas except linearity and reported in Tab. \ref{tab_off}.
 
A similar approach was performed with $Y$ and $Z$ but due to the distribution of the 3 stops about $X$ (every $\frac{2\pi}{3}$\, rad) at each end of the test-mass, the position under gravity when relying on $Y$ or $Z$ is not as precise as for $X$. That's why in Tab. \ref {tab_off} the figures for $Y$ and $Z$ can be biased  by a few microns. In that case $d_1$ and $d_2$ are determined as upper or lower limit values to verify that the free motion is sufficient to enable a good operation of the servo control with margins. The offcentering along $Y$ and $Z$ in Table \ref{tab_off} are mainly determined by the metrology of the cylinders once integrated by considering capacitive effects as negligible. 
On ground and in flight measurements are compatible at 3$\sigma$ and shows the difficulty to predict the flight centring to better than 8$\mu$m with ground metrology alone mainly because it is not possible to levitate the test-mass (see section \ref{sect_sagel}). Error bars are also difficult to determine with metrology from different sources and methods. 

The centring $\vv{O_1O_2}$ estimated in flight are also presented in Table \ref{tab_off} by extracting from the accelerometric measurement the effect of the Earth's gravity gradient \cite{hardycqg6}. On ground several errors cumulate: 4$\mu$m error on the test-mass position lying on the stops (deduced from 18 mechanical measurements at 1$\mu$m accuracy each) and 2$\mu$m metrology error on the geometry. In flight, the error comes from the dispersion of the different estimations made with the Earth's gravity field knowledge and the accelerometric measurements \cite{hardycqg6}, without any temperature dependency correction.

\begin{table}
\caption{\label{tab_off} Evaluation of the SU test-masses relative centring ($^*$errors do not take into account capacitive border effects).}
\begin{indented}
\item[]\begin{tabular}{@{}lrrr}
\br
  & X & Y & Z \cr
\mr
On-ground estimation&&& \cr
SU-EP & $24\pm{}6\,\mu$m & $0\pm{}4\,\mu$m$^*$ & $-6\pm{}4\,\mu$m$^*$ \cr
SU-REF & $-28\pm{}6\,\mu$m & $4\pm{}4\,\mu$m$^*$ & $0\pm{}4\,\mu$m$^*$ \cr
\mr
In-flight estimation&&& \cr
SU-EP & $20.1\pm{}0.1\,\mu$m & $-8.0\pm{}0.8\,\mu$m & $-5.7\pm{}0.1\,\mu$m \cr
SU-REF & $-35.5\pm{}0.3\,\mu$m & $5.6\pm{}0.4\,\mu$m & $5.7\pm{}0.3\,\mu$m \cr
\br
\end{tabular}
\end{indented}
\end{table}

\subsubsection {Core metrology: test-mass alignment.}
The test-mass alignments limit the projection of undesired signals on the X axis. There are specifications on the value and its knowledge for a potential correction. The mean alignment of the test-mass with the star sensor has to be known to better than $10^{-3}$\,rad to limit the error of the Earth's gravity gradient projection on the $X$ axis during the data process. The orthogonality of the test-mass axes has also to be set to better than $10^{-4}$\,rad to limit cross-axis projection. This requirement is completed by the test-mass relative alignment of $1.5\times{}10^{-3}$\,rad calibrated in-orbit to an accuracy of $10^{-4}$\,rad to reject the common mode accelerations. This last requirement was verified in flight and showed some good margins \cite{hardycqg6}. 

The sensors alignment with respect to the SUMI reference frame is obtained on the three faces of an optical cube glued on the SUMI (Fig. \ref{fig_core}). The budget is established considering defects such as flatness of the SUMI, parallelism of the base plate and presented in Tab. \ref{tab_align} and shows good agreement with the requirement of $1.5\times{}10^{-3}$\,rad.

\begin{table}
\caption{\label{tab_align} On ground budget error and estimation accuracy of the SU alignment with respect to the SUMI Optical Cube reference.}
\begin{indented}
\item[]\begin{tabular}{@{}lrr}
\br
  & Around $Y$ or $Z$ & Around $X$ \cr
  & rad & rad \cr
\mr
SU-EP && \cr
Assembly budget & $10^{-4}$ & $4.6\times{}10^{-4}$ \cr
Accuracy  & $0.8\times{}10^{-4}$ & $4.5\times{}10^{-4}$ \cr
\mr
SU-REF && \cr
Assembly budget & $1.2\times{}10^{-4}$ & $7.1\times{}10^{-4}$ \cr
Accuracy  & $10^{-4}$ & $7.1\times{}10^{-4}$ \cr
\mr
Requirement &  $15\times{}10^{-4}$ & $15\times{}10^{-4}$ \cr

\br
\end{tabular}
\end{indented}
\end{table}

\subsection{T-SAGE instrument electronics on-ground verification} \label{sect:Elecground}

\subsubsection {Capacitive sensor.}
As a sensor, the capacitive position detector is characterised by an equivalent input noise spectrum, an equivalent input bias, a gain, a dynamic range and non-linearity parameters. Both gain and bias parameters have their own thermal sensitivity. The main sources of bias are the dissymmetries originating from the differential transformers, the stray capacitance difference of the coaxial cables connecting each electrode pair to the primary winding and the two superimposed capacitors allowing for the action feedback on each opposite electrode. A defect of symmetry of the differential transformers, thus of the inductance $\delta L$ introduces a detector output voltage proportional to $V_d$ and $\delta L$. The differential transformers are manually wound and finely tuned in order to get the primary coils as symmetric as possible. The relative symmetry $\delta L / L$ achievement is better than $10^{-5}$.

The superimposed capacitors associated to each electrode pair has to exhibit the smallest current leakage by choosing high capacitance values and getting the same values. Each superimposed capacitor is carefully sorted in order to reach the best matching values before implementation.

In order to minimise the bias, at the last stage of the FEEU production, the detector channel bias is adjusted by the addition of a stray capacitance of a few picofarads in one of the two branches. By doing so, the maximum bias obtained was limited to 50\,mV, corresponding to a capacitance misbalance of $6\times{}10^{-4}$\,pF.

The bias detector thermal sensitivity error source is found to be in the differential transformer and extensive tests have been conducted in order to select the best techniques in term of coil shielding, coil potting and coil packaging to minimise the thermal effects.

\begin{figure}
\begin{center}
\includegraphics[width=0.5\textwidth]{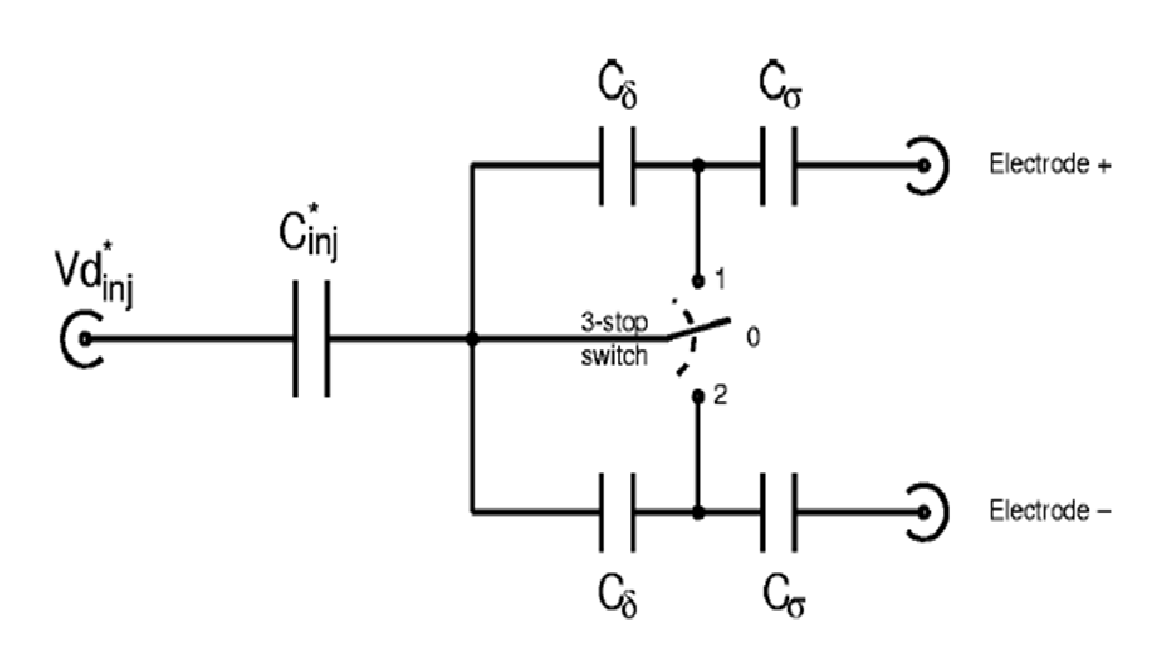}
\caption{A baseline for the capacitances of the test bench set to : $C_{inj}=1$\,pF, $C_\sigma=1.2$\,pF and $C_\delta=27$\,pF. The misbalance equals to $0.005$\,pF when the switch is set on position 1 or 2, short-circuiting one of the 2 $C_\delta$. The ``0 position'' shall lead to a capacitance balance. Since it is $C_\sigma$ that drives the misbalance figure, the tolerance has been set to a maximum of $7$\,fF between the two capacitances $C_\sigma$. 1\% accuracy capacitances are implemented for $C_{inj}$ and $C_\delta$.}
\label{fig_test}       
\end{center}
\end{figure}

The measurement of the electromechanical transducer characteristics is not straightforward \cite{petrucha10}. Indeed the mechanical core cannot operate under ground gravity. Thus a mechanical core simulator has been used as a specific item of the test equipment that allows to operate and to test the capacitive detector individually. It simulates the test-mass / electrodes capacitance thanks to ceramic capacitors. A schematic is provided in Fig. \ref{fig_test}. The motion of the test-mass is simulated by modifying the balance of the test-mass / electrodes capacitors. This bi-level command is performed thanks to a switch. The amplitude of the so-called motion is controlled by the input 100 kHz voltage $V_{d_{inj}}$. Amplitude modulation with $V_d$ as carrier permits to simulate a low-frequency displacement signal. The capacitances are measured with an accurate capacitance bridge (Andeen-Hagerling\textregistered{}\,AH 2500) in order to fully characterise the SU simulator with femtofarad accuracy.

To obtain the sensitivity measurement, the detector output is recorded for the three positions of the SU simulator, and then redone with permutation of the electrode cables. This provided six measurements used to determine the detector and simulator properties simultaneously from the equation: $V=G_{\rm det} (\Delta C_i +b_s)+b_{\rm det}$, 
where $\Delta C_{i=0,1,2}$ is the approximately 0, +1, or -1 pF capacitance difference of the simulator, depending on the switch position $(i=0,1,2)$, $b_s$ is its bias, and the entire term between the parentheses becomes negative when the simulator is connected in the opposite side.  This method eliminates the bias induced by the simulator itself. Hence the detector gain $G_{det}$ and bias $b_{det}$ are determined.

To demonstrate the detector linearity a change in proof mass position is simulated by changing the amplitude of the reference voltage input $V_{d_{inj}}$ to the SU simulator while maintaining the demodulator reference at $5$\,Vrms. The amplitude is always positive to stay in phase with the demodulator reference ($V_{d0}$) signal and the sign change is obtained by using the two positions of the simulator. The input variation created by changing the $V_{d_{inj}}$ amplitude is converted to a corresponding variation of capacitance $\Delta C = \Delta C_0 V_{d_{inj}} /  V_{d0}$, where $V_{d0}$ is the 5\,V$_{\rm rms}$ level input to the detector demodulator.

\begin{figure}
\begin{center}
\includegraphics[width=1\textwidth]{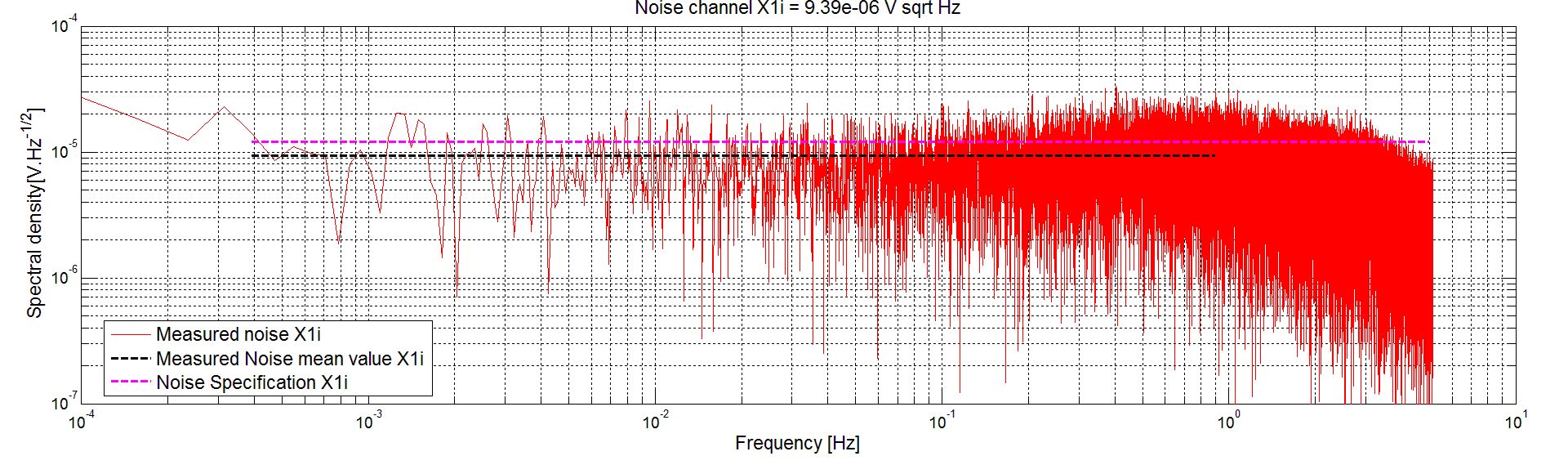}
\caption{Typical detector spectrum noise with an example of channel X1i (X channel, inner test-mass).}
\label{fig_spectX1i}       
\end{center}
\end{figure}

The detector noise is measured by grounding the input of the SU simulator. The major noise sources are the transformer and the first amplifier. The frequency domain of interest is from very low frequency down to $10^{-4}$\,Hz to approximately $100$\,Hz, corresponding to the control loop bandwidth. This requires long time measure in a stable thermal enclosure. Fig. \ref{fig_spectX1i} shows a typical noise spectrum after 7 hours of data recording. The data analysis required a down-sampling from 1027\,Hz to 10\,Hz associated with a type averaging FIR filter, which explains the decrease of the spectrum with respect to frequency above 5\,Hz.

The thermal sensitivity measurement especially on the detector bias requires a particular procedure due to the small amount of variation expected. The classical method consisting in measuring $N$ values of bias at $N$ temperatures (the FEEU box placed in a thermal chamber) is not efficient here since the measurement points are under the noise level. To measure such a small fluctuation, a different method has been applied for MICROSCOPE electronics consisting in the application of low frequency sinusoidal thermal fluctuation to the electronic box at $2\times{}10^{-4}$\,Hz  and to retrieve the thermal bias sensitivity by frequency analysis. 

Tab. \ref{tab_detector} summarises the main measurement results obtained on MICROSCOPE's FEEU flight models capacitive sensors. Taking into account the theoretical physical gains in Tab. \ref{tab_gain}, the measurements show a position sensor resolution better than $4\times{}10^{-11}$m\,Hz$^{-1/2}$ for any linear test-mass degree of freedom and better than $8.5\times{}10^{-9}$rad\,Hz$^{-1/2}$ for any angular test-mass degree of freedom. Concerning the bias, the impact in terms of position sensing is lower than 0.08$\mu$m and in terms of angular sensing lower than 35$\mu$rad.

\begin{table}
\caption{\label{tab_detector} Laboratory measured resolution, bias, thermal sensitivity and gain, of the capacitive sensors at 20\,\textdegree{}C.}
\begin{indented}
\item[]\begin{tabular}{@{}lrrrrrr}
\br
Capacitive sensor's resolution & $X$ & $Y$ & $Z$ & $\phi$ & $\theta$ & $\psi$  \cr
at $10^{-2}$\,Hz [$\mu$V\,Hz$^{-1/2}$] &&&&&& \cr
\mr
IS1 SU-REF & 8.1 & 3.8 & 3.7 & 9.3 & 3.8 & 3.7 \cr
IS2 SU-REF & 5.4 & 2.1 & 2.1 & 9.6 & 2.1 & 2.1 \cr
IS1 SU-EP   & 12 & 3.1 & 3.1 & 12 & 3.1 & 3.1 \cr
IS2 SU-EP   & 5.6 &1.9 & 1.9 & 5.5 & 1.9 & 1.9 \cr
\br
Bias [mV] & $X$ & $Y$ & $Z$ & $\phi$ & $\theta$ & $\psi$  \cr
\mr
IS1 SU-REF & -6 & -7 & -1 & 14 & 2 & -3 \cr
IS2 SU-REF & 14 & -1 & -1 & -46 & 0 &1  \cr
IS1 SU-EP   & 14 & -1 & -9 & 49 & 1 & -4 \cr
IS2 SU-EP   & 21& -2 & -5 & 30 & 1 & 1 \cr
\br
Thermal sensitivity [$\mu$V/K]& $X$ & $Y$ & $Z$ & $\phi$ & $\theta$ & $\psi$  \cr
\mr
IS1 SU-REF & 64.7 & 10.1 & 14.3 & 36 & 1.7 & -2.1 \cr
IS2 SU-REF & 16.7 & 12.0 & 25.4 & 130.9 & 2.7 &12.2  \cr
IS1 SU-EP   & 65.9 & 23.3 & 10.7 & 37.7 & 10 & -0.6 \cr
IS2 SU-EP   & 16.3 & 8.9 & 14 & 193.3 & 1.2 & -0.5 \cr
\br
Capacitive sensor's gain [V/pF]& $X$ & $Y$ & $Z$ & $\phi$ & $\theta$ & $\psi$  \cr
\mr
IS1 SU-REF & 82.5 & 16.9 & 17.3 & 82.2 & 16.9 & 17.3 \cr
IS2 SU-REF & 40.6 & 5.0 & 5.2 & 84.5 & 5.0 & 5.2  \cr
IS1 SU-EP   & 81.2 & 16.0 & 16.0 & 81.0 & 16.0 & 16.1 \cr
IS2 SU-EP   & 39.3 & 5 & 5 & 85.0 & 5.0 & 5.0 \cr
\br
\end{tabular}
\end{indented}
\end{table}

\subsubsection {Electrostatic actuation channel test.}
An actuation channel is composed of a pair of electrodes with, associated to each one, a Drive Voltage Amplifier (DVA) and a DAC that commands the DVA.

The DVA is a low noise feedback amplifier capable of delivering voltages in the range of about $-40$\,V to $+40$\,V when powered with $\pm{}48$\,V power lines. It consists of 4 stages:
\begin {itemize}
\item A 16-bit DAC that feeds the return branch of the accelerometer loop with the control voltage provided by the PID in the ICU. Its performance in terms of noise and linearity is a critical item for the resolution of the instrument;
\item A second stage based on an operational amplifier which matches impedance and isolates the amplifier input from noise; the internal feedback, although not theoretically necessary, prevents the DVA from auto-oscillations;
\item A third stage which delivers the output voltage across a capacitive load.
\end {itemize}

As part of the feedback branch of the accelerometer, the DVAs are one of the major contributors to the accelerometer performance: they impose noise and linearity limits. 
Since the measurement bandwidth starts at $0.1$\,mHz, the main criterion for the selection of an operational amplifier technology for this function is its behaviour at this bound. While previous models using OP97 exhibit a typical slope of $1/f$ noise in the measurement bandwidth, chopper-stabilized amplifiers are free of this effect. Then the equivalent noise of the amplifier is that of the first stage.

The linearity and thermal sensitivity of DVAs have been measured to assess the stability of the scale factor. The resistors involved in the transfer function have a thermal stability better than $0.05$\,ppm/\textdegree{}C.

Tab. \ref{tab_dva} gathers the main results obtained on the DVA electronics of the X axis. All gains of Electrode + and Electrode - have been matched to better than $0.015$\% for X axis and $0.02$\% for all the other axes which have very similar gains. For what concern the bias, all others axes exhibit a maximum bias of 1.7\,mV and a maximum bias sensitivity of 3.5$\mu$V/\textdegree{}C.

\begin{table}
\caption{\label{tab_dva} Bias and thermal sensitivity of the actuator electronics (DVA) at electrodes $X_+$ and $X_-$ (which control the translation along the $X$ axis) as measured in the laboratory before flight at 20\textdegree{}C.}
\begin{indented}
\item[]\begin{tabular}{@{}lcc}
\br
  & Electrode $X_+$ & Electrode $X_-$ \cr
\mr
\multicolumn{3}{l}{Gain of DVA :} \cr
IS1-SUREF  & $-15.93472$ & $-15.93264$ \cr
IS2-SUREF  & $-15.93419$ & $-15.93323$ \cr
IS1-SUEP  & $-15.93285$ & $-15.93053$ \cr
IS2-SUEP  & $-15.93227$ & $-15.93926$ \cr
\mr
\multicolumn{3}{l}{Noise (any test-mass) : $0.4\mu$V\,Hz$^{-1/2}$} \cr
\mr
\multicolumn{3}{l}{Bias of DVA ($\mu$V) :} \cr
IS1-SUREF  & $157$ & $168$ \cr
IS2-SUREF  & $118$ & $146$ \cr
IS1-SUEP  & $147$ & $225$ \cr
IS2-SUEP  & $256$ & $254$ \cr
\mr
\multicolumn{3}{l}{Thermal sensitivity $\mu$V/K :} \cr
IS1-SUREF  & $0.3$ & $1.0$ \cr
IS2-SUREF  & $-0.8$ & $0.9$ \cr
IS1-SUEP  & $-2.0$ & $-1.1$ \cr
IS2-SUEP  & $0.4$ & $-0.9$ \cr

\br
\end{tabular}
\end{indented}
\end{table}

\subsubsection {Reference voltage test.}

The $V_p$ voltage is obtained by the average of two very stable voltage references in parallel. In order to guarantee a very stable scale factor, the $V_p$ function thermal stability has been specified to $3$\,ppm/\textdegree{}C. $V_p$ production and characteristics relies mainly on the quality of the DC source used. As the requirement is more stringent than what can be found on the data sheet for space qualified components (ceramic case), the same reference in plastic case with much better performance has been sorted to select the best pair of components for each $V_p$ function. $V_p$ has been measured in flight as $V_p = 5.003\pm{}0.013$V with respect to the FEEU zero voltage (through the 1\,Hz housekeeping) and can be compared to ground characterisations in Table \ref{tab_refvolt}.

The 100\,kHz $V_d$ detection voltage amplitude modulation noise is measured through a rectified diode circuit in order to characterise the low frequencies behaviour. 

Tab. \ref{tab_refvolt} gathers the main results obtained on ground.

\begin{table}
\caption{\label{tab_refvolt}On-ground characterisation of reference voltages: $V_p$ and $V_d$ functions at 20\,\textdegree{}C.}
\begin{indented}
\item[]\begin{tabular}{@{}lrrrrr}
\br
 & IS1-SUREF & IS2-SUREF & IS1-SUEP & IS2-SUEP  \cr
\br
\multicolumn{5}{l}{$V_p$}\cr
Value [V] & 4.9973 & 4.9983 & 4.9997 &4.9990 \cr
Broadband noise [$\mu$V\,Hz$^{-1/2}$] & 0.16&0.16 &0.22 &0.22\cr
Corner frequency [Hz] &2&2&2&2\cr
Thermal sensitivity [ppm/\textdegree{}C] & -2.1& -1.2 & -1.0 & -0.9\cr
\br
 \multicolumn{5}{l}{$V_d$}\cr
Value [V${_rms}$] & 5.0282 & 5.0152 & 5.0288 &5.0188 \cr
Frequency [kHz] & 99.91 & 99.92 & 99.86 & 99.86\cr
Broadband noise [$\mu$V\,Hz$^{-1/2}$] & 1.5&1.5&1.5 &1.5\cr
Corner frequency [Hz] &2&2&2&2\cr
Thermal sensitivity [ppm/\textdegree{}C] & 50& 50 & 49 & 59\cr
\br
\end{tabular}
\end{indented}
\end{table}

\subsubsection {Read-out channel test.}
The ``off-the-loop'' or read-out channel takes a measurement at electrode level by doing a difference between electrode $X_{+}$ and electrode $X_{-}$. This design was meant to help reject the DVA noise present in the loop. This measurement is captured by an ADC. As for the DVA function, a very accurate attenuation factor conditioning the signal before the digital conversion, has been obtained by the use of ultra-high precision and stability resistor in order to use this channel for science purpose. The obtained thermal sensitivity of the bias was lower than $1.8\,\mu$V/\textdegree{}C and of the gain lower than 7.9\,ppm/\textdegree{}C. The four channels show a bias lower than 1mV and a noise lower than $1.6\,\mu$V\,Hz$^{-1/2}$. 

In spite of this outstanding performance on ground, the SUEP-IS2 $X$ output exhibited in flight an acceleration higher than expected, identified as a bias due to a stronger gold wire stiffness. This led to the saturation of the read-out channel on this test-mass. The read-out channel was afterward only used for technology purposes. Furthermore the noise is dominated by this gold wire parasitic disturbance and thus minimises the interest of using this channel eventually exhibiting the same performance as the ``in-the-loop'' channel.

\subsubsection {Digital control validation.}
The servo-loop correctors are implemented in the rad-hard 32/40-bit IEEE floating point DSP TSC21020F. The control loop operation cycle was defined at 1027\,Hz leading to a CPU load of 77\% typical and 95\% maximum. Calculation accuracy was a key point throughout the development, computation errors introduced by the digitisation inevitably appears during the operations of quantification and rounding. Both are difficult to detect and therefore had needed careful tests and measurements.

Moreover in such a DSP, the variables which intervene in the different computations are memorised using the IEEE 754 floating point standard with 32 bits and the computations (additions or multiplications) are performed using a 40-bit format. For some critical operations, the error analysis showed that the variables have to be stored in 40\,bits for the rounding errors to be negligible. For the read-out channel, the global measurement error due to the 32/40-bit conversion format is dominated by the amplification of error during the different filtering processes. This issue has been solved by the use of a new computation method which minimises the error propagation. It had been shown that replacing the regular filter by a null gain filter improves the results significantly.

For on-ground test, the digital control has been very convenient to validate the ``in-the-loop'' electronic functions by implementing appropriate control law or by simply implementing unitary control law.

In flight, the digital control has permitted to optimise the servo control by patching new control law parameters, and even to implement a new functionality: the possibility to store 8 seconds of actuation data at 1027\,Hz (instead of 4\,Hz in the regular telemetry) for investigation purpose.


\section{T-SAGE instrument validation} \label{sect_sagel}

The development of the payload has benefited from the experience acquired with four previous space accelerometer missions: ASTRE with COLUMBIA shuttle, STAR \cite{touboul99b} with CHAMP mission, SuperSTAR with GRACE mission \cite{flury08}, and the 6 accelerometers of GOCE mission \cite{marque10}. Nevertheless drastic modifications are considered because of a new constraint: the operating ranges of the accelerometer in its Full or High Resolution Modes (FRM and HRM) are not compatible with ground operations because of the intensity of the gravity. The needed electrostatic forces are out of access. Nevertheless these ranges remain reachable with the accelerometric environment in the Bremen free-fall drop tower \cite{selig10} in certain conditions. 

We have followed the classical development flow of engineering model (EM), qualification model (QM) and flight models (FM) for both the SU and the electronics FEEU and ICU. However, the SU development was more complex due to operational constraints explained above. 

First, SU prototype and pre-qualification models have been developed to assess the vibration resistance and thermal behaviour, then an SU engineering model has been realised and tested with light silica test-masses in order to be levitated on ground. This step allowed us to operate the digital control of the loops and to establish the control laws for the future qualification model tested in the free-fall tower. 

\begin{figure}
\begin{center}
\includegraphics[width=0.25\textwidth]{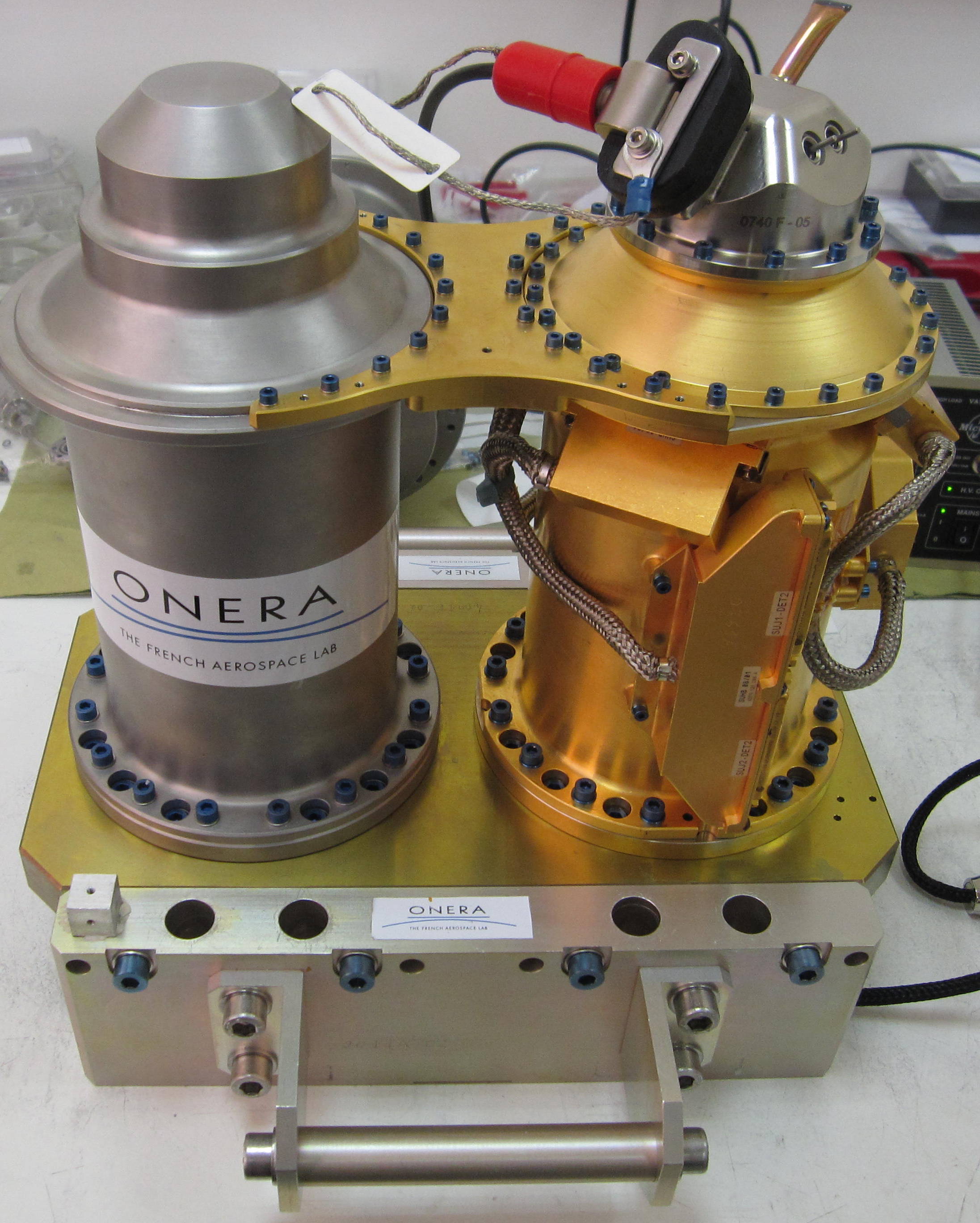}
\includegraphics[width=0.3\textwidth]{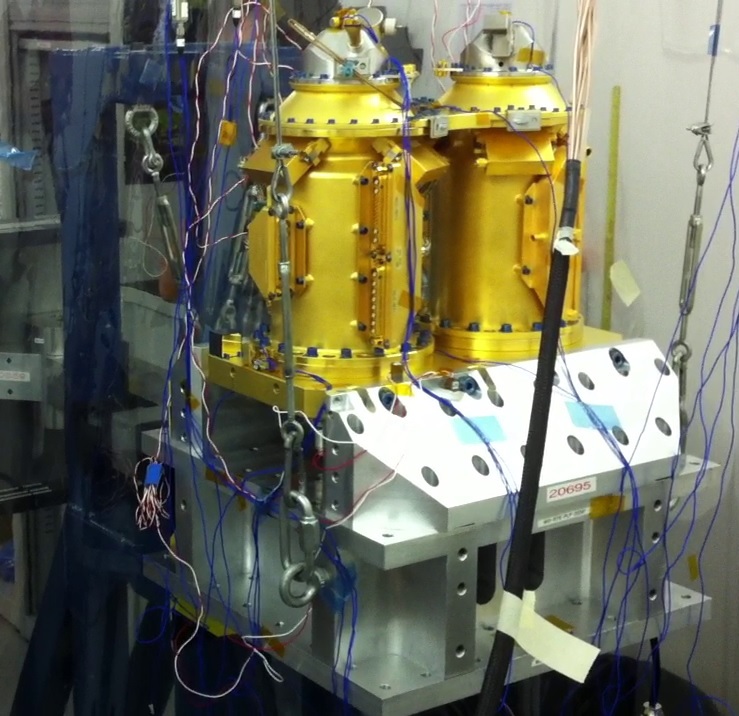}
\includegraphics[width=0.35\textwidth]{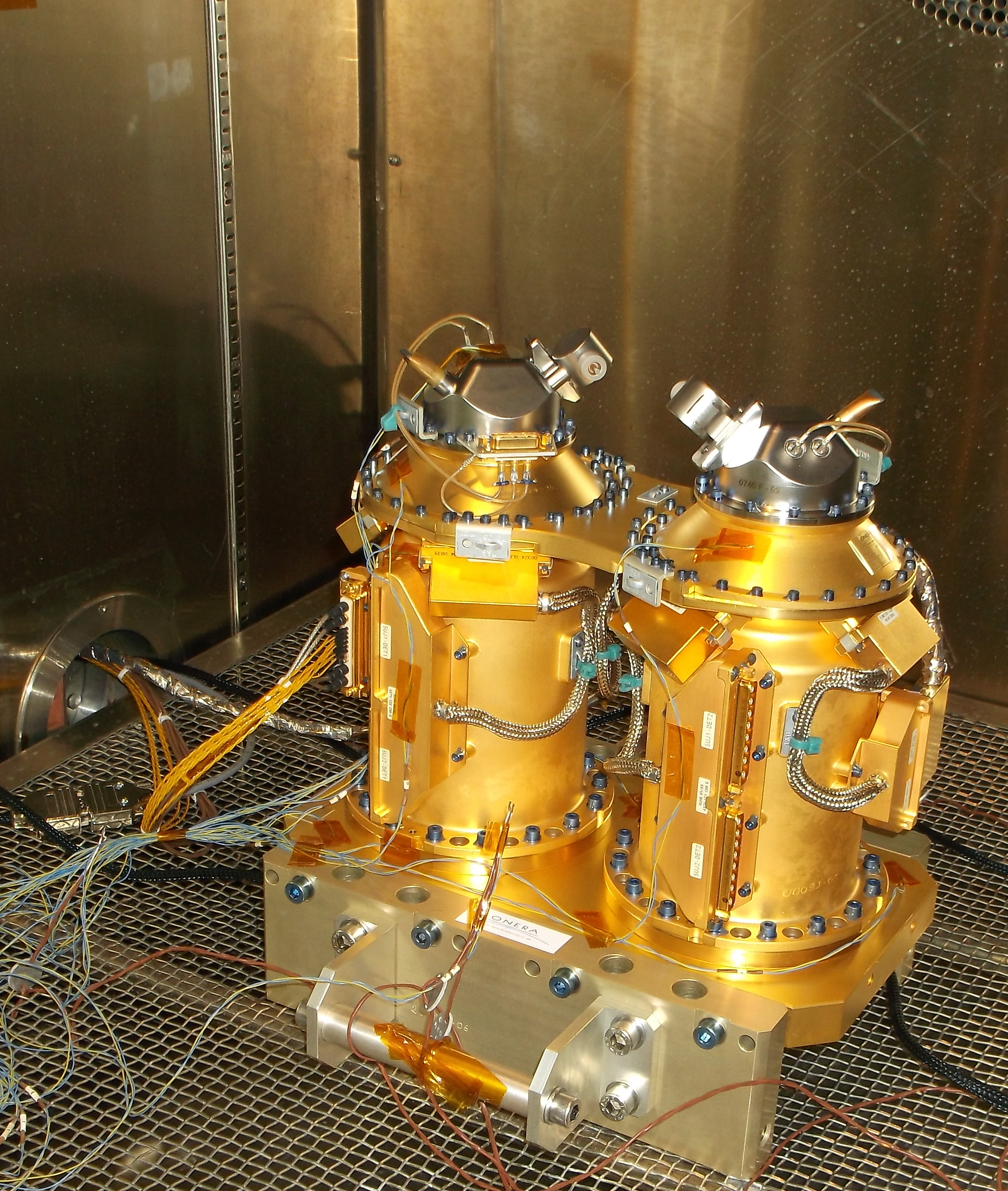}
\caption{Left: P5.6 vibration model for the gold wire resistance. Middle: QM model under shock qualification tests. Right: FM model under thermal acceptance tests.}
\label{fig_qualif}       
\end{center}
\end{figure}

Then two particular vibration test models, representative of the flight model for the mechanical aspects only, have been produced, assembled and vibrated. The first test model was dedicated to the validation of the test-mass blocking system when submitted to the vibration environments. It also allowed us to check the free motion of test-masses after vibration and switch on of the release system. The second test model was fully representative of the flight model from a mechanical point of view, with fewer engravings and less cabling since its purpose was to establish the integration processes and to test the general behaviour of the core mechanics under vibration conditions. It does not feature the vacuum pump and was not functional (left on Fig. \ref{fig_qualif}).

Later, the Qualification Model (QM) (middle on Fig. \ref{fig_qualif}) has been integrated as fully identical to the flight model except for the test-masses in wolfram alloy test-masses (Inermet\textregistered{}). Inermet\textregistered{}\, density is 17.6, 10\% lower than the PtRh density. The QM has been used for final vibrations, shocks, thermal and EMC testing and was operated during free-fall tests in the ZARM drop tower. 

Finally, the two Flight Models have undergone campaigns with limited vibration and thermal tests (Fig. \ref{fig_qualif}). The development of the instrument concluded on a final drop test in the ZARM free-fall tower prior to the integration into the satellite and the launch.

\subsection {SU vibration tests}
The payload was delivered to CNES in October 2014. The first qualification process specified a 9\,g$_{rms}$ load for the random vibrations in the frequency band [20\,Hz-2000\,Hz]. Unlike previous accelerometers developed by ONERA, T-SAGE possesses a blocking device in order to handle the 0.4\,kg to 1.4\,kg test-masses safely during the launch phase (test-masses are surrounded by silica cylinders and might have been damaged by moving masses). However, this blocking device has a dynamic motion at eigen frequencies (400\,Hz, 600\,Hz and 800\,Hz) of a few hundreds of microns that led to a fatigue of the $7\,\mu$m gold wire. Two years of development were necessary to understand the fatigue process and to correct it. 

In cooperation with CNES, a specific damping device was implemented at payload interface with the satellite that reduces to 3.5\,g$_{rms}$ the levels of vibration and thus the amplitude of the motion of the gold wire. The integration process was also improved to guarantee a repeatability of the gold wire gluing and a better resistance. Before delivery, the qualification model successfully sustained 5 times the vibration duration that would be applied to the flight model from acceptance tests to in-orbit operation. This endurance test, quite out of the generic qualification process, helped to get confidence on the resistance of such a sensitive payload.

The levels of vibrations were less critical to sustain for the electronics, and no endurance tests were necessary.

\subsection{Thermal assessment}
\begin{figure}
\begin{center}
\includegraphics[width=0.55\textwidth]{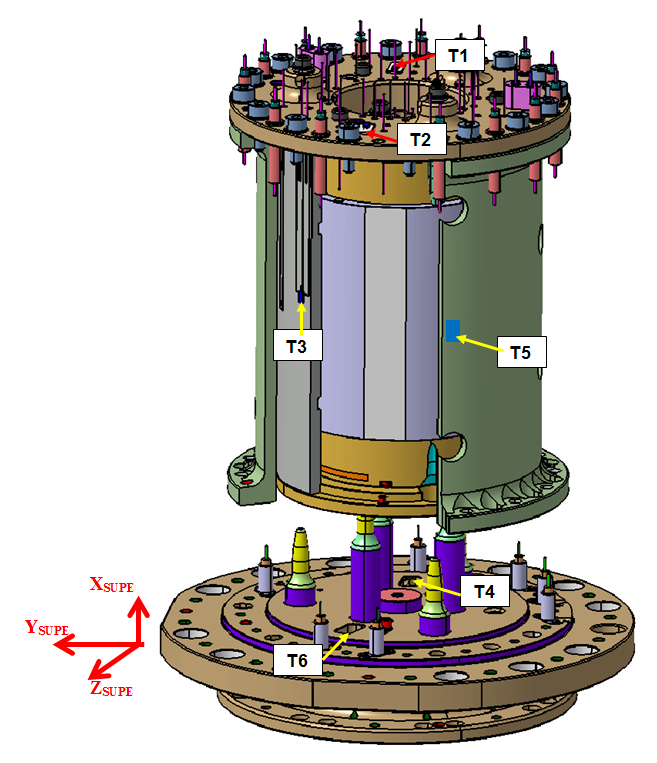}
\caption{Position of the SU internal thermal probes.}
\label{fig_probe}       
\end{center}
\end{figure}

The two Sensor Units on their interface plate (SUMI) and the two FEEU are integrated in the payload case with highly stabilised thermal environment. The temperature variation is the main error source. Six temperature probes (Fig. \ref{fig_probe}) have been integrated into each SU, and five temperature probes have been integrated into each FEEU in order to carefully monitor the evolution of the thermal environment within the payload case.  Particular tests have been performed at satellite level, to evaluate the thermal performance of the payload case for which a passive stabilisation was chosen.  The SU temperature variations were specified at $f_{\rm{EP}}$ to better than 1\,mK at SU level and 5\,mK at FEEU level. A dedicated model was developed with representative thermal characteristics and power dissipation in order to evaluate the way the different thermal power sources can affect the payload at $f_{\rm{EP}}$. The test was done in CNES Toulouse premises. The test result analysis has allowed readjusting the numerical thermal model of the payload case with accuracy better than 15\% \cite{hardycqg6}.

\subsection{Functional assessment: drop tower test}

\subsubsection{Facility and instrument configuration.}
A unique facility in Europe, the drop tower in ZARM Institute provides conditions for experiments under microgravity. The height of the tower limits the “free-fall” experiment period to 4.74\,s for drops, extended to 9.5\,s since December 2012 with catapult shots \cite{kampen06}.
For the MICROSCOPE mission, 25 campaigns with a total of 71 falls have been conducted in order to assess the dynamic behaviour of T-SAGE twelve digital servo-loops: 49 drops with the “free-flyer” \cite{liorzou14} and 22 drops with the catapult. Previously the drop capsule in “free-flyer” configuration was settled with a SuperSTAR accelerometer (engineering model of GRACE mission) with some 100 drops. The flight models were tested with only 8 drops to limit the stress.
QM campaign drops are presented and discussed in Ref. \cite{liorzou14}. In the following, the catapult shot results obtained with the QM model are shown and the very last test, performed in June 2014 of the flight model is presented.

Due to the very short free-fall duration and the small range of the accelerometer, the electronics configuration had to be boosted to 100\,V instead of the nominal flight supply of 48\,V in order to increase the applied voltage capabilities and thus the electrostatic forces for the test-mass acquisition. The DC voltage applied to the mass is also changed from 41.4\,V to 89.2\,V. With this drop tower configuration, the full scale range of $X$ measurement for the smallest of the 4 test-masses is of $2.7\times{}10^{-5}$\,m\,s$^{-2}$ versus $1.6\times{}10^{-6}$\,m\,s$^{-2}$ in flight HRM mode.

The accelerometers and their electronics are accommodated in a cylindrical capsule (2\,m high and 80\,cm in diameter). Its weight is about 300\,kg with its electronics, its on-board data acquisition computer and the experiment payload. At the bottom of the capsule, a 50\,cm long cone stabilises it during the deceleration inside the 8 meters high braking tank filled with fine graded polystyrol. The deceleration process lasts about 0.2\,s and the maximum deceleration rate does not exceed 400\,m\,s$^{-2}$ ($\approx$40\,g) and can be applied without damage to the instrument. For the free-fall, the $X$ axis is always horizontal, the $Y$ axis is aligned along gravity (the same axis as the cylindrical capsule) and the $Z$ axis is also horizontal. The residual drag of the capsule is thus measured along the $Y$ axis.  Optimisation of the capsule balance is a key point in the experiment preparation: without any prior balancing of the capsule, the centre of gravity is not located on the capsule axis of symmetry anymore, also aligned with the axis joining the two sensor units along $Y$. The test-masses positions versus the centre of gravity of the capsule are better than 1\,mm along the radial axes.

A final pressure of about 10\,Pa is achieved inside the drop tube in order to minimise the drag effect at 200\,$\mu$m\,s$^{-2}$. A particular care is also taken to any motion inside the capsule that may generate small vibrations which may be measured by the accelerometers: for instance, cables are carefully fixed and the on-board computer fans are switched off. 

\subsubsection{Drop results.}

Since the instrument is configured in a coarse acquisition mode because of the short free fall time, its ultimate performances are out of reach. However, such drops are the only way to test the servo control dynamics. Their objective is to demonstrate that the sensor operates as expected, especially that the test-mass is controlled at the zero position with desaturated acceleration measurements along all degrees of freedom. 

\begin{figure}
\begin{center}
\includegraphics[width=0.99\textwidth]{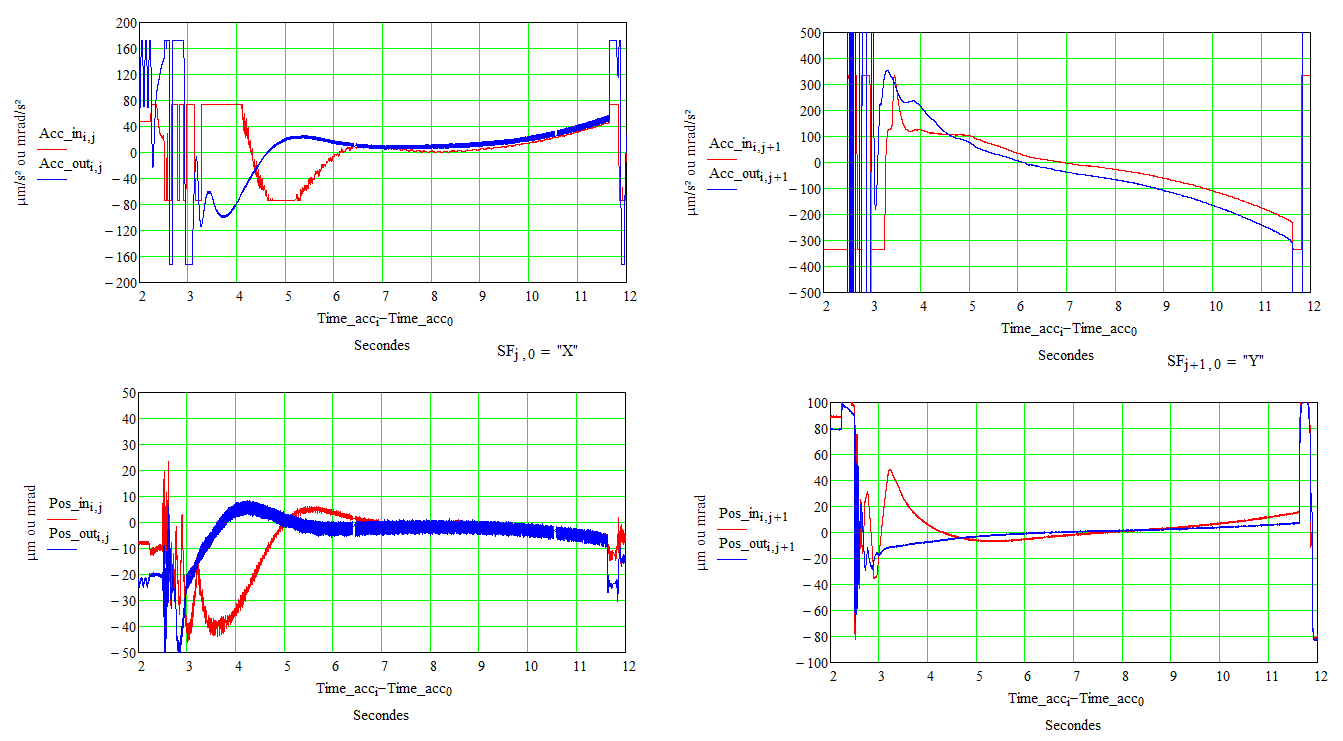}
\caption{Measurements during the last catapult test of SU-REF and SU-EP flight models. On the left, the $X$ axis measurements. On the right, the $Y$ axis measurements. The upper graphs show the acceleration of inner (red) and outer (blue) test-masses. The lower graphs show the position of the test-masses.}
\label{fig_drop}       
\end{center}
\end{figure}

Fig. \ref{fig_drop} illustrates a typical catapult test with the flight models. Before the capsule release, under 1\,$g$, the test-masses lay on the stops. Along the $Y$ axis (vertical, direction of the fall), they are resting at a distance of about $+90\,\mu$m from the centre of the electrode set (see section \ref{centering}) with $+Y$ direction pointing to the ground. The instrument is switched on before the release of the capsule, so when under microgravity, the acquisition of the mass electrostatic suspension is automatically performed. The capsule is catapulted to the top of the ``free-fall'' tower. As soon as the level of the capsule acceleration is within the accelerometer full range, the mass position is controlled to the centre of the electrode set and then the acceleration measurement is desaturated (Fig. \ref{fig_drop}).

In a standard drop, the drag due to the residual air in the tower is continuously increasing from the time of the capsule release to the end of the drop. During a catapult shot the drag decreases to zero at the apex of the trajectory i.e. at the very precise moment when the velocity is also zero. In Fig. \ref{fig_drop}, the bias of the instrument can be estimated  at the apex, however this determination is not precise because the test-masses are not yet controlled along and about all directions particularly along $X$.
With the radial electrodes, the control of the test-masses is sufficiently stiff to get the convergence of the PID. The measured acceleration provided by each accelerometer can be considered as the sum of the external acceleration and the accelerometer bias. Thus the simple difference of the measurements provides an estimate of the differential bias in a coarse mode ($V_p=100\,V$).

\subsection{Last instrument ``good health test''}

As the accelerometers cannot operate on ground, good health tests have been established in order to assess the integrity of the instrument during the integration on the satellite, the satellite environmental test and after transportation to the launch pad site. It consists in verifying as exhaustively as possible the functionality of the SU, FEEU and ICUME knowing that the instrument is in launch configuration: test-masses locked on their stops.

The main functionalities verifications are obtained through dedicated housekeeping data and through the scientific data themselves contained in the telemetries:
\begin {itemize}
\item At SU level: the gold wire health is checked thanks to a command electrically disconnecting the stops. The capacitive detectors must show then the position of the test-masses locked on the stops if the $V_d$ voltage is well applied by the wire. 
\item A force sensor measures the blocking mechanism efficiency: a total $1600$\,N is measured during locked phases and less than $200$\,N (the sensor bias) when the test-masses are free.
\item At FEEU level: the capacitive detector channels provide repeatable measurements with different ``expected" values with respect to the two values of $V_d$. This validates the electronics, the SU/FEEU harness and the SU core integrity at once. 
\item At ICU level: the digital control loop through the open loop response is verified as well as the data production, the command control, and the power supply.
\end {itemize}


\section{Conclusion} \label{sect_ccl}

T-SAGE inherits from previous developments of electrostatic accelerometers \cite{touboul04, touboul99, touboul99b}. The major improvement is the use of a digital controller for the test-mass servo loop with a large capability of configuration change or parameterisation in order to make this instrument as user friendly as a laboratory experiment. Another difference with previous developments was the impossibility to levitate the test-masses on ground which led to a development strongly relying on free-fall tests for the servo loop adjusting phase and for the qualification verifications.

The other key points are the test-mass shape, the cylindrical test-masses and the gold-coated electrode sets needed new procedures of machining and metrology. 
Finally it was the first time, a locking device system is used for the test-masses. Previously, the test-masses did not weigh more than 0.3\,kg and mechanical studies and experience showed that they could be free even during the launch vibration phase. With the 1.3\,kg test-mass, a locking device was mandatory. New developments were necessary to define a locking device that could be safely released in orbit, with no damage on the surface of the test-mass and that could not induce sticking forces at the release or disturbing forces during flight. 
The electrodes around the test-mass and all others surfaces that could load an electrical charge have been carefully designed to limit any electrostatic disturbances on the test-mass motion. In particular the mechanical stops have multilayer coatings to differentiate areas of different voltages in front of electrodes or of the test-mass.

The confidence in the final performance of the instrument has relied on the verification of each individual specification in the mechanics (dimension, position, mass…), in the environment (level of vacuum, temperature, magnetic characteristics…) and in the electronics (noise, offset, gain, thermal stability…). The performance cannot be measured on ground which is why a particular care has been taken to establish each contributor to the detailed error budget that contributes to the mission performance budget as detailed in \cite{rodriguescqg1}.


\ack

The authors express their gratitude to all the different services involved in the mission partners and in particular CNES, the French space agency in charge of the satellite. This work is based on observations made with the T-SAGE instrument, installed on the CNES-ESA-ONERA-CNRS-OCA-DLR-ZARM MICROSCOPE mission. ONERA authors’ work is financially supported by CNES and ONERA fundings.
Authors from OCA, Observatoire de la C\^ote d’Azur, have been supported by OCA, CNRS, the French National Center for Scientific Research, and CNES. ZARM authors’ work is supported by the German Space Agency of DLR with funds of the BMWi (FKZ 50 OY 1305) and by the Deutsche Forschungsgemeinschaft DFG (LA 905/12-1). The authors would like to thank the Physikalisch-Technische Bundesanstalt (PTB), in Brunswick, Germany, for their contribution to the development of the test-masses with funds of CNES and DLR.

\section*{References}
\bibliographystyle{iopart-num}
\bibliography{biblimscope}

\end{document}